\documentclass{emulateapj}

\shorttitle{LAL observations}
\shortauthors{M. Luna et al.}

\begin{document}

\title{Observations and Implications of Large-Amplitude Longitudinal Oscillations in a Solar Filament}

\author{M. Luna, \altaffilmark{1,2}, K. Knizhnik\altaffilmark{3,4},  K. Muglach\altaffilmark{5,4}, J. Karpen\altaffilmark{4}, H. Gilbert\altaffilmark{4}, T.A. Kucera\altaffilmark{4}, \& V. Uritsky\altaffilmark{4,6}}

\altaffiltext{1}{Instituto de Astrof{\'{\i}}sica de Canarias, E-38200 La Laguna, Tenerife, Spain}
\altaffiltext{2}{Universidad de La Laguna, Dept. Astrof{\'{\i}}sica, E-38206 La Laguna, Tenerife, Spain}
\altaffiltext{3}{Department of Physics \& Astronomy, The Johns Hopkins University, Maryland, USA}
\altaffiltext{4}{NASA GSFC, Maryland, USA}
\altaffiltext{5}{ARTEP, Inc., Maryland, USA}
\altaffiltext{6}{Catholic University of America at NASA GSFC, Greenbelt, MD 20771, USA}

\begin{abstract}
On 20 August 2010 an energetic disturbance triggered large-amplitude longitudinal oscillations in a nearby filament. The triggering mechanism appears to be episodic jets connecting the energetic event with the filament threads. In the present work we analyze this periodic motion in a large fraction of the filament to characterize the underlying physics of the oscillation as well as the filament properties. The results support our previous theoretical conclusions that the restoring force of large-amplitude longitudinal oscillations is solar gravity, and the damping mechanism is the ongoing accumulation of mass onto the oscillating threads. Based on our previous work, we used the fitted parameters to determine the magnitude and radius of curvature of the dipped magnetic field along the filament, as well as the mass accretion rate onto the filament threads. These derived properties are nearly uniform along the filament, indicating a remarkable degree of cohesiveness throughout the filament channel. Moreover, the estimated mass accretion rate implies that the footpoint heating responsible for the thread formation, according to the thermal nonequilibrium model, agrees with previous coronal heating estimates.  We estimate the magnitude of the energy released in the nearby event by studying the dynamic response of the filament threads, and discuss the implications of our study for filament structure and heating.  
\end{abstract}

\section{Introduction}\label{sec:intro}

High-cadence H$\alpha$ observations of large-amplitude longitudinal (LAL) oscillations in solar filaments were first reported by \citet{jing2003}. These oscillations consist of rapid motions of the plasma along the filament, with displacements comparable to the filament length. Since then, a few more events have been identified \citep{jing2006,vrsnak2007,li2012,zhang2012}. In all of the observed oscillations the period ranges from 0.7 to 2.7 hours, with velocity amplitudes from 30 to $100~\mathrm{km~s^{-1}}$. The accelerations are considerable, in many cases comparable to the solar gravitational acceleration. In addition, the oscillations are always triggered by a small energetic event close to the filament.

Explaining the LAL oscillations is very challenging because the restoring force responsible for the huge acceleration must be very strong. The energy of the oscillation is also enormous, because the filament is massive  and large velocities are generated. However, the motions damp quickly in a few periods, implying a very effective damping mechanism. Several models have been proposed to explain the restoring force and damping mechanism of the LAL oscillations \citep[see review by][]{tripathi2009}, but most do not successfully describe the thread motions. Recently, we studied the oscillations of threads forming the basic components of a filament \citep{luna2012a} in a 3D sheared arcade \citep{devore2005}. In this model, the threads reside in large-scale dips on highly sheared field lines within the overall magnetic structure. We found that the restoring force is mainly gravity, and the pressure forces are small  \citep{luna2012b}. This result has been confirmed with numerical simulations by \citet{li2012} and \citet{zhang2013a}.  We also studied analytically the normal modes of plasma on a dipped field line, and found that the oscillation generally involves the superposition of two components:  gravity-driven modes and slow modes associated with pressure gradients \citep{luna2012c}. However, for typical filament plasma properties the gravity-driven modes dominate. This type of oscillation resembles the motion of a gravity-driven pendulum, where the frequency depends only on the solar gravity and the field-line dip curvature. We estimated the minimum value of the magnetic field at the dips and found agreement with previous estimates and observed values. Additionally, this study revealed a new method for measuring the radius of curvature of the filament dips. These studies demonstrated that the LAL oscillations are strongly related to the filament-channel geometry. 

We identified the damping of the LAL oscillations as a natural consequence of the thermal nonequilibrium process most likely responsible for the formation and maintenance of the cool filament threads \citep{luna2012b}. In this process, the localized footpoint heating produces chromospheric evaporation and subsequent collapse of the evaporated mass into cool condensations in the corona \citep[e.g.,][]{antiochos1999, karpen2006}. As long as the heating remains steady, the threads continuously accrete mass and grow in length. Non-adiabatic effects (i.e., thermal conduction and optically thin radiation) also contribute weakly to the damping \citep{luna2012b, zhang2013a}. The damping times obtained with this model generally agree with the observed values. 

Because LAL oscillations in filaments are observed to occur after small energetic events nearby, we speculated that local heating and/or flows from the energetic event triggers and drives oscillations in filament threads magnetically linked to the event site \citep{luna2012b}. \citet{zhang2013a} modeled the effects of two types of perturbations on a filament thread -- impulsive heating at one leg of the loop and impulsive momentum deposition -- and determined that both can cause oscillations. 

In this work we study a LAL filament oscillation, its trigger, and subsequent damping observed on 20 August 2010 by the Atmospheric Imaging Assembly (AIA) instrument on the Solar Dynamics Observatory (SDO) \citep{Lemen2012a}. We also utilize co-temporal magnetograms from the Helioseismic and Magnetic Imager (HMI) instrument on SDO \citep{scherrer2012}, to provide the overall magnetic context and local connectivity between the filament and the triggering event. The oscillation properties yield the geometry of the magnetic field supporting the cool plasma and the mass accretion rate. This information leads to fundamental conclusions about the filament structure, as well as the dynamics and energetics of the triggering event. Some preliminary results of this work were shown in \citet{Knizhnik2014}. 

In \S\ref{sec:observation} we describe the data and sequence of events, while \S\ref{sec:trigger} discusses the event that triggered the oscillations. In \S\ref{sec:analysis} our methodology for measuring the filament oscillation and deriving key parameters is described.  \S\ref{sec:results} presents the results and examines their implications for the oscillation mechanism, mass accretion, and energization. Based on these results, we describe the most likely filament structure in \S\ref{sec:structure}. Our findings and conclusions are summarized in \S\ref{sec:conclusion}.

\section{Observations and event overview}\label{sec:observation}

AIA/SDO and HMI/SDO observe the entire solar disk with regular, high temporal cadence. AIA obtains images in seven EUV filters at 12-s cadence, and in two UV filters at 24-s cadence. HMI takes polarized full-disk images of the Stokes $I$ and $V$ parameters, and the data pipeline yields line-of-sight (LOS) magnetograms, Dopplergrams, and intensity images at a 45-s cadence.

The events considered in this investigation occurred in a large filament channel centered in the southern-hemisphere active region (AR) NOAA 11100, which was studied in detail for a different time period by \citet{Wang2011a} and \citet{Wang2013a}. For an overview of the AR and the filament, including H$\alpha$ images and a HMI LOS magnetogram, we refer the reader to Figure 1 of \citet{Wang2013a}.
 
\begin{figure*}
\centering\includegraphics[width=16cm]{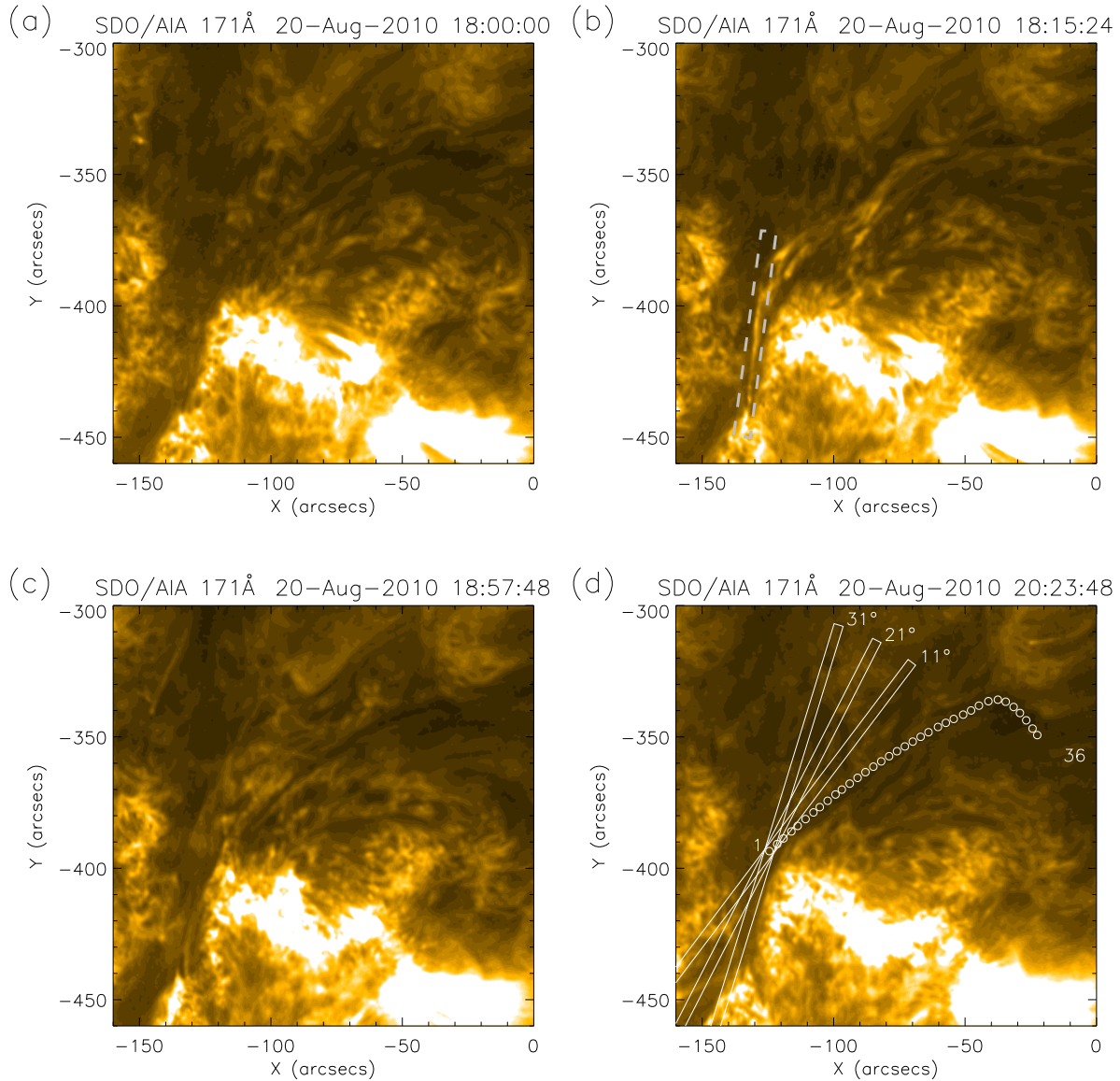}
\caption{Time sequence of 171 \AA\ images containing the analyzed filament and jets. (a) Just before the impulsive event. b) Fully developed jet flows visible in emission. The dashed box outlines the slit used to study the jet in \S \ref{sec:trigger}. c) The SE part of the filament is clearly excited and displaced toward the north. d) The oscillating filament. White circles indicate the 36 positions along the filament selected for analysis; for clarity only positions 1 and 36 are numbered. White boxes outline three examples of slits at position 1 as described in \S \ref{sec:analysis}.  A movie of the AIA 171 \AA\ images of the region shown is available in the online journal. 
\label{fig:refmap}}
\end{figure*}

On 20 August 2010, the decayed active region was located east of central meridian in the southern hemisphere at about $-27^\circ $ latitude. A long filament cuts diagonally through the AR, and consists of two  segments in the southeastern and northwestern regions of the AR. Our analysis is focused on the northwestern part of the filament outlined by the $160'' \times 160''$ region shown in Figure \ref{fig:refmap} in the AIA 171 \AA\ channel at 4 times between 18:00 and 20:23 UT. Therefore, all further references in this paper to ``the filament" apply to the northwest segment only. The analyzed temporal sequence started at 15:00 UT August 20 and ended at 05:00 UT August 21. SDO takes science data almost continuously, interrupted only by brief eclipses and calibration sequences, which were taken into account in the analysis. The AIA and HMI images were transformed to the same spatial scale and co-aligned with the SolarSoft routine ``aia\_prep.pro'' and the mapping routines developed by D. Zarro.

The filament is visible as a dark band in H$\alpha$ and in most of the AIA EUV filters (171 \AA , 193 \AA , 211 \AA\ and 304 \AA). The dark band corresponds to the region where the cool plasma resides, characterized by increased absorption of background emission and an emission deficit in coronal lines (see Figure \ref{fig:refmap}). In addition, the edges of the dark band clearly exhibit emission in these lines. \citet{schmieder2004a,schmieder2004} and \citet{Parenti2012} studied this absorption/emission pattern in filaments, and found that the absorption pattern resembles the H$\alpha$ morphology while the bright emission comes from the prominence-corona transition region (PCTR). In fact our thermal nonequilibrium model predicted this bright emission from the PCTR at both ends of filament threads \citep{luna2012a}.

A movie of AIA 171 \AA\ images of the region shown in Figure \ref{fig:refmap} is available in the online journal, showing the sequence of events during the 3 hours when most of the interesting activity occurs. At the beginning of the observation interval (15:00 UT) the motions inside the filament channel resemble the well-known counterstreaming \citep[see][]{zirker1998,alexander2013}, with different parts of the structure moving out of phase. Figure \ref{fig:refmap}(a) shows a representative example of this initial stage, although the counterstreaming displacements are not evident in this still image. Around 18:10 UT a narrow collimated flow of plasma appears near the south end of the filament (see \S \ref{sec:trigger}). In Figure \ref{fig:refmap}(b) this jet is clearly visible inside the overplotted dashed box and bright plasma is seen over the entire filament. By 19:00 UT the outflow from the jet source has ceased and the filament oscillates with a very coherent, large-scale motion. In this first hour after the jet onset, the flow appears to push the existing filament threads toward the northwest, particularly in the southeastern portion of the filament (Figure \ref{fig:refmap}(c)). The filament continues oscillating (Figure \ref{fig:refmap}(d)) until the end of the temporal sequence.
 
After 19:00 UT the filament oscillations are evident in all AIA filters. As is obvious from the movie, however, not all sections of the filament oscillate: the central portion remains at rest throughout the event, while the east and west segments oscillate (see \S\ref{sec:analysis}). The oscillation damps quickly at first but continues throughout the 14-hour observation. At 21:20 UT an eruption occurs outside our studied region but close to the NW end of the oscillating filament, centered at $\{x \sim 60'',y\sim -420''\}$. The erupting filament appears to come from the SW extension of the same large-scale filament channel under study, but only the closest oscillating filament threads (i.e., slits 33-34) are affected significantly. For maximum clarity we focus on the oscillations produced by the jet, before the nearby eruption.

\section{Trigger Analysis}\label{sec:trigger}

Around 18:10 UT very collimated plasma flows start to emanate from $\{x \sim -135'', y \sim -440''\}$, ejecting plasma toward the SE end of the filament situated $\sim70''$ ($\sim$51~Mm) away from the source (see Figure \ref{fig:refmap} and accompanying movie). This jet is clearly visible in emission in several AIA EUV filters (171 \AA, 193 \AA, 131 \AA, 304 \AA, 211 \AA, and 94 \AA), indicating that the multi-thermal flow includes very hot plasma. At the same time, a secondary jet appears to come from a position north of the primary jet source ($\{x \sim -130'', y \sim -420''\}$). The presence of filamentary cool plasma between these source locations makes it difficult to determine whether the secondary jet really is independent, or whether the primary jet is merely obscured by the filament between the two positions. In either case, these streams merge rapidly and proceed as one highly structured flow along the filament channel. A detailed analysis of the jet is beyond the scope of this work, and would not affect the present analysis and interpretation of the oscillations. Therefore, in subsequent discussions we treat the jet as a single entity originating at the southernmost location.

\begin{figure}
\centering\includegraphics[width=9cm]{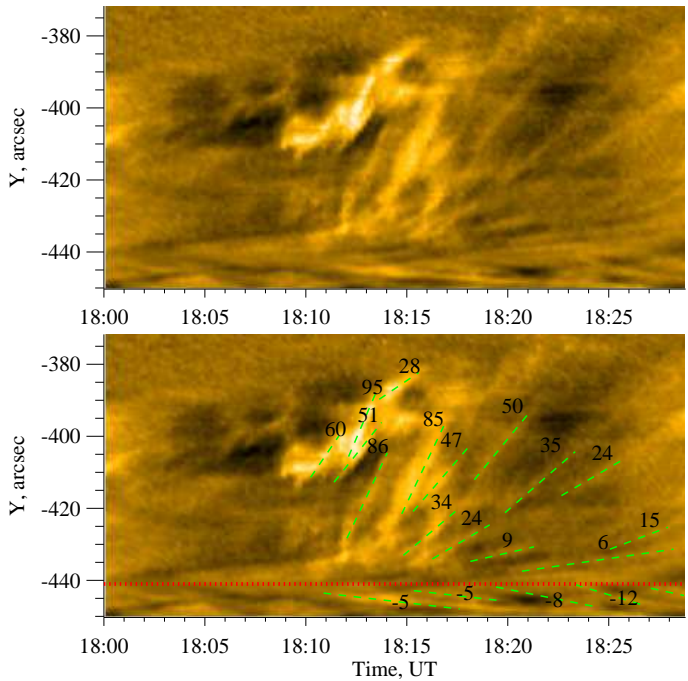}
\caption{Time-distance diagrams of the episodic initiation jet. a) AIA 171 \AA\ intensity extracted from the dashed box indicated in Fig. \ref{fig:refmap}(b) on the y-axis; time is given on the $x$-axis (starting at 18:00 UT). b) Same as a) with linear fits of several propagating fronts associated with individual ejection episodes overplotted in green. The red dotted line gives the approximate y-location of the jet origin, with plasma emission propagating either northward or southward from this position along the slit. The numbers give the estimated speed from these fits in $\mathrm{km~s^{-1}}$ \label{fig:frontspeeds}}
\end{figure}

Initially the outflow is narrowly focused and linear (see movie). Therefore, we placed an artificial slit over the image (indicated by the dashed box in Figure \ref{fig:refmap}(b)) and extracted the intensities along this slit as a function of time between 18:00 UT and 18:30 UT. The intensity signals were logarithmized and subject to a third-order polynomial detrending, which allowed us to improve the signal-to-noise ratio. We avoided temporal smoothing to preserve the original time resolution of the AIA data \citep{uritsky2013}. The resulting time-distance diagram is shown in Figure \ref{fig:frontspeeds}, where the bottom panel displays the same image as the top panel with linear best fits to the diagonal intensity structures overplotted as green dashed lines. The derived outflow speeds are given for each fitted structure. The flow is episodic with a few minutes between successive ejections. The flow speed reaches $\sim 95~ \mathrm{km~s^{-1}}$ in the first episode, and decreases significantly in subsequent episodes. These values are consistent with the speeds of different jets in the same filament channel analyzed by \citet{Wang2013a}, although those jets did not produce filament oscillations.

Most emission structures in Figure \ref{fig:frontspeeds} are tilted upward, indicating that the flow is moving northward, although at the bottom of the Figure a few flows are moving southward. The red dotted line marks the line of demarcation between these cotemporal flows. Several of the jets described in \citet{Wang2013a} are also clearly bi-directional. Applying the same analysis method to the AIA 193 \AA, 335 \AA\ and 304 \AA\ images gives very similar time-distance diagrams as Figure \ref{fig:frontspeeds} and the same values for the speeds of the fronts. Therefore we conclude that the flow speed is independent of temperature, and that we are measuring the bulk speed of the plasma ejected from the source. These speeds are projected onto the image plane, however, so the low southward values are most likely from flows that are primarily vertical. Furthermore, the apparent acceleration of the northward flows might indicate a smooth transition from vertical to horizontal field-line geometry, as we expect from the magnetic structure of the sheared arcade model. In the absence of Doppler information derived from emission line profiles, however, we can only speculate about the true velocities.  

The most compelling interpretation of these observations is that the bi-directional jets are reconnection outflows originating in the low corona. If the reconnection occurred in the chromosphere or photosphere, the downward jet would not be seen. \citet{Wang2013a} found evidence for flux cancellation associated with their bi-directional jets observed near the same filament channel, providing support for reconnection as the driver. However, we studied the HMI LOS magnetograms at our jet source regions and found no unambiguous evidence for changes in magnetic flux density, neither for mutual flux cancellation nor flux emergence. The photospheric flux density values are very close to the observed noise level in the filament channel, and fluctuate constantly on short time scales (from a few minutes down to the 45-s cadence of the magnetograms). 

The initial jet flows are visible in emission in all AIA EUV filters, but after the impulsive phase when the threads are perturbed by the hot flows, there is not enough hot plasma to produce emission in the 94 \AA\ filtergram. Later in the initiation event the jet path becomes less collimated and more turbulent, with some of the bright features reversing direction (see Figure \ref{fig:refmap} movie). Further analysis and interpretation of the spatial and energetic relationship between the jet and the filament threads is discussed in \S \ref{sec:init}, \S \ref{sec:energy}, and \S \ref{sec:conclusion}.

\section{Methodology: oscillation analysis}\label{sec:analysis}

The LAL oscillation event is clearly visible in H$\alpha$ and the AIA/SDO filters 171 \AA, 193 \AA, and 131 \AA, but harder to discern in the 304 \AA\ passband. The ground-based H$\alpha$ images are useful for context, but they have lower spatial resolution and cadence than the AIA/SDO data. Therefore we identified and parameterized the oscillations from the EUV images, placing slits centered along the filament at different angles with respect to it. We have chosen this method because it is not possible to identify and track individual threads or features from the images. We determined the boundaries of the filament by eye, following the dark band in the 171 \AA\ images, and placed the center of each slit at the midpoint of the dark band. As shown in Figure \ref{fig:refmap}(d), we selected 36 slits whose centers are separated by 6 pixels (3.6 arcsecs) along the total (projected) filament length of 126 arcsecs (92 Mm). The slits are numbered from 1 to 36 starting from the southeast end of the filament and ending at the northwest end. The filament is composed of two quasi-linear segments: the eastern part is the longest, containing the first 28 slits, while the short western part is tilted $66^\circ$ with respect to the other and contains the last 8 slits. Each slit is 3 pixels (1.8 arcsecs) wide and 300 pixels (180 arcsecs) long. The intensity was averaged over the 3 transverse pixels and the resulting intensity distribution along the slit was plotted as a function of time, as the primary tool for analysis of the oscillations (Figures \ref{fig:bestslits1}-\ref{fig:bestslits3}). Note that we utilized the full 12-s cadence of AIA data.

The best EUV passband for this analysis is 171 \AA\, because the dark and bright regions have the highest contrast with very sharp profiles. The filament slice seen in each 171 \AA\ slit appears as a dark band surrounded by two bright bands. This pattern allows us to clearly identify the motions of the filament threads.

\begin{figure*}[!ht]
\centering\includegraphics[width=18cm]{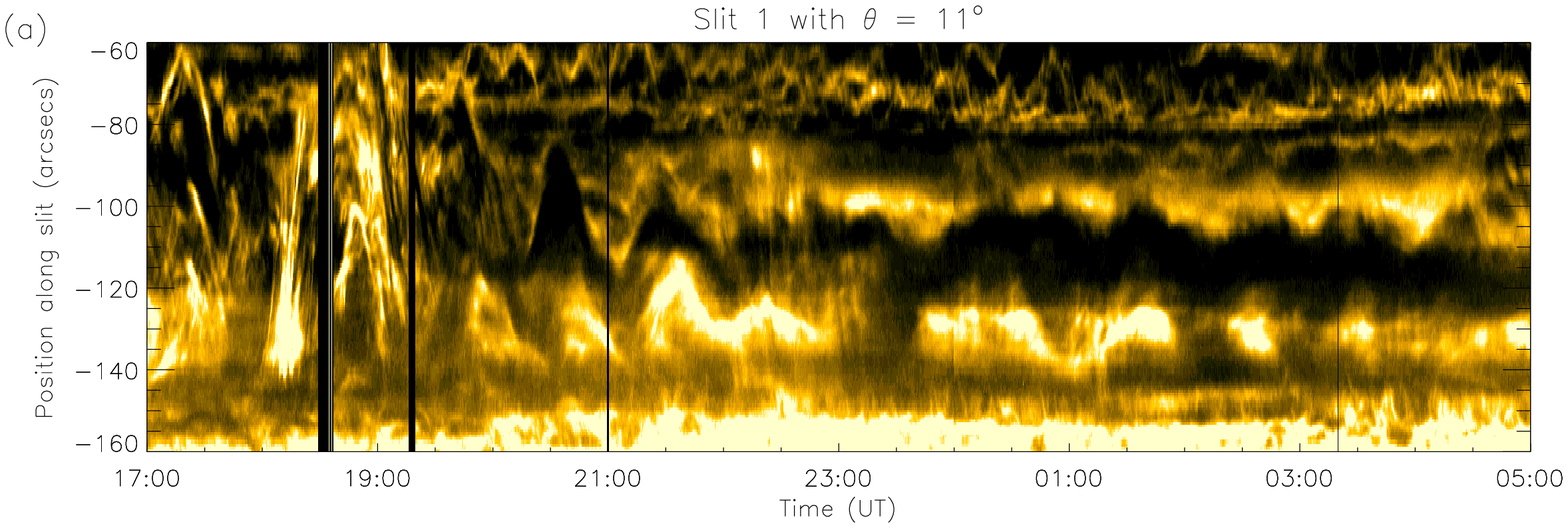}\\
\centering\includegraphics[width=18cm]{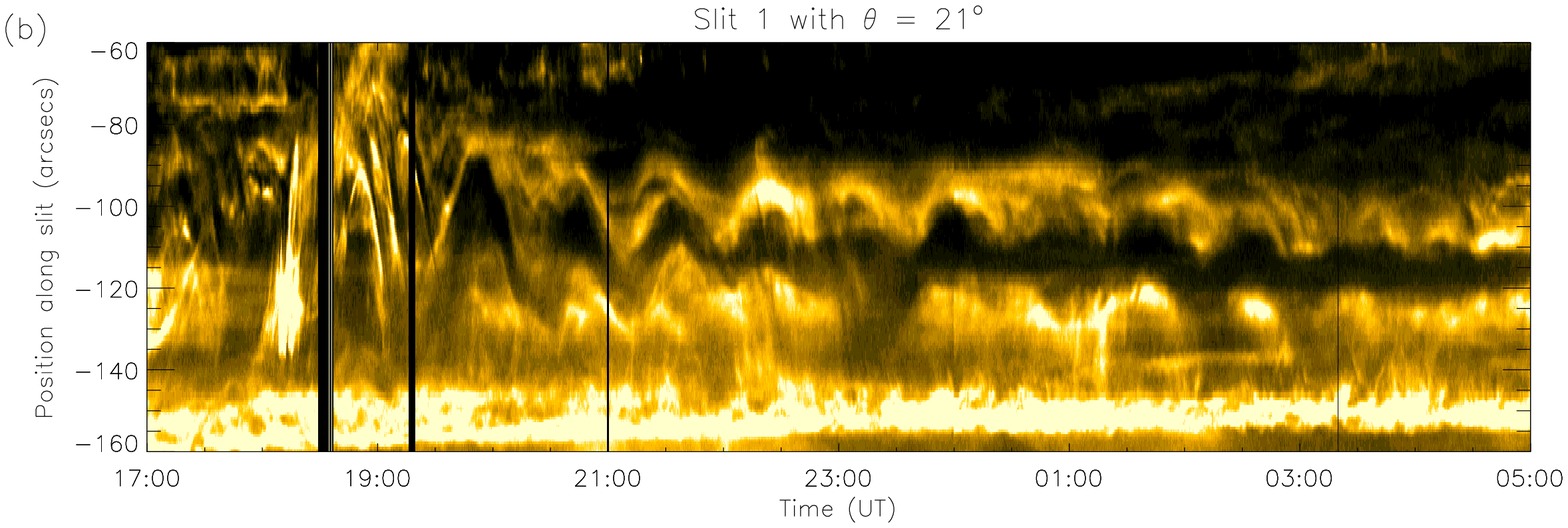}\\
\centering\includegraphics[width=18cm]{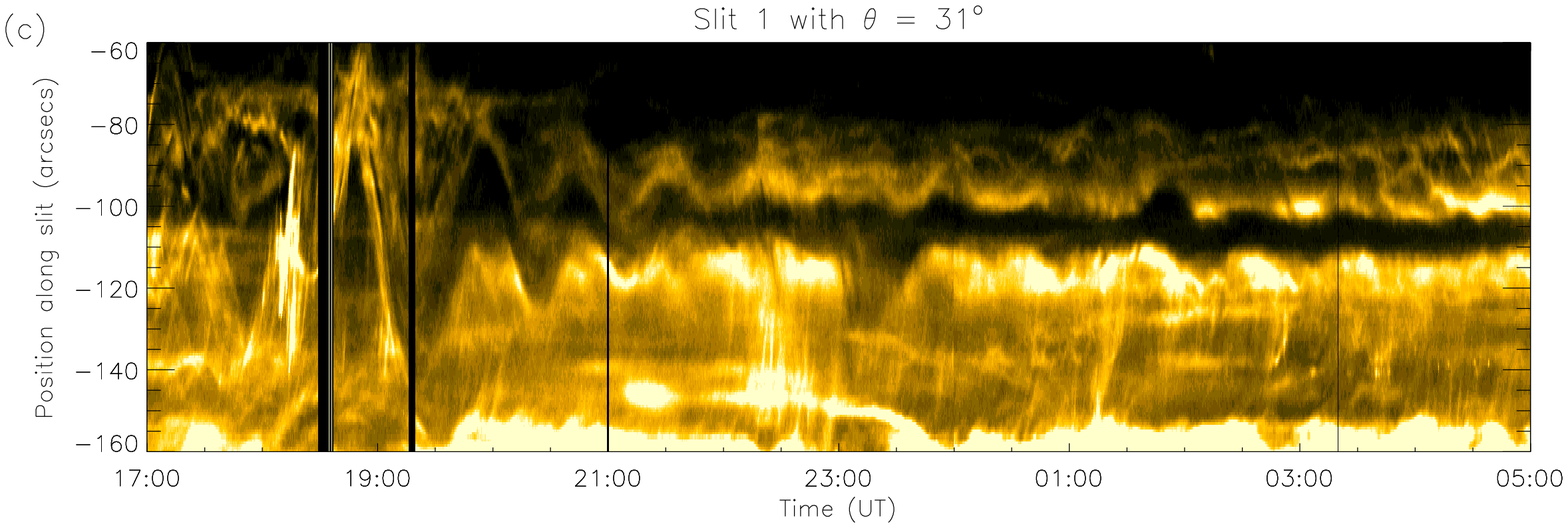}
\caption{Time-distance diagrams from SDO/AIA 171 \AA\ images in slit position 1 on the filament with angles (a) $11^\circ$, (b) $21^\circ$, and (c) $31^\circ$. The slits used to generate these three diagrams are shown on Fig. \ref{fig:refmap}(d). Data gaps are due to eclipses and calibration sequences. The plot begins on 20 August and ends 21 August    2010. 
\label{fig:anglescan}}
\end{figure*}

\begin{figure*}[!ht]
\centering\includegraphics[width=18cm]{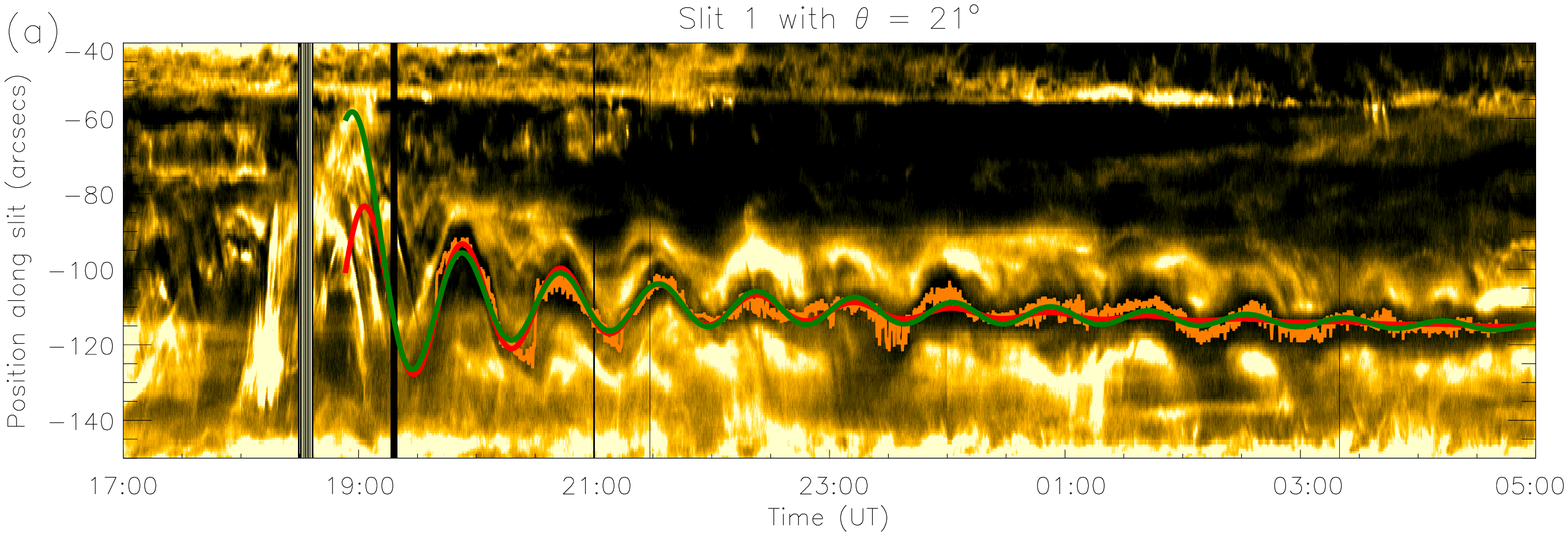}\\
\vspace{0.1cm}\centering\includegraphics[width=18cm]{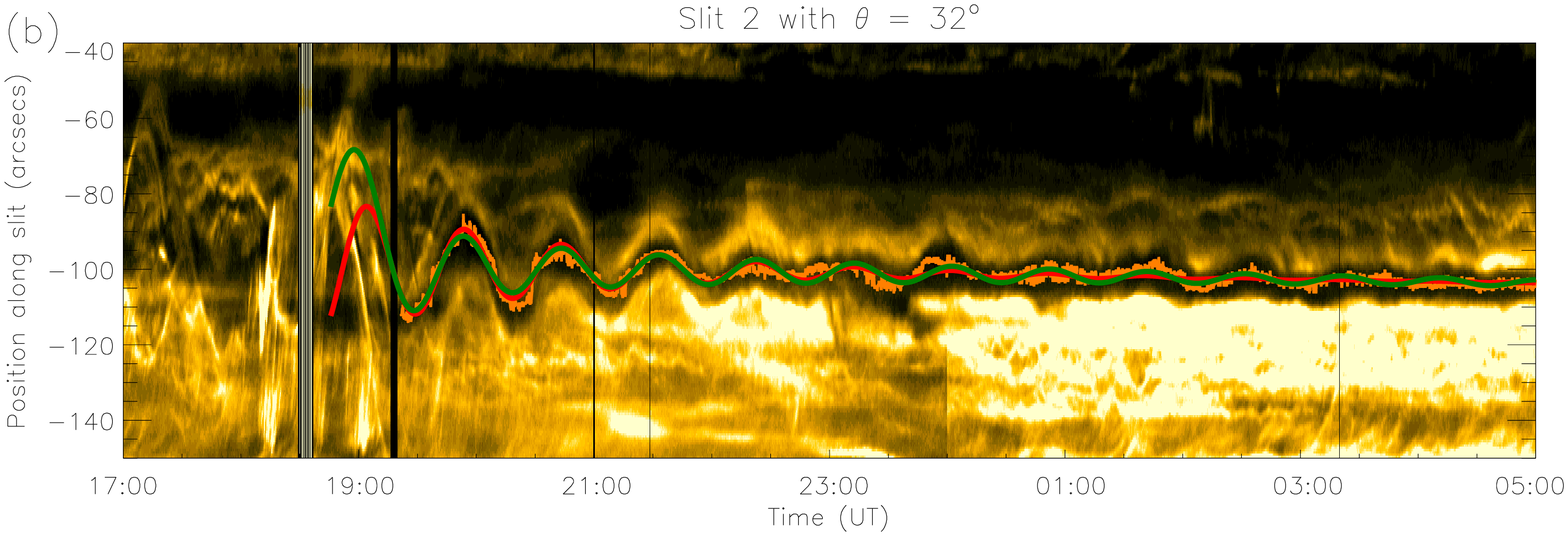}\\
\vspace{0.1cm}\centering\includegraphics[width=18cm]{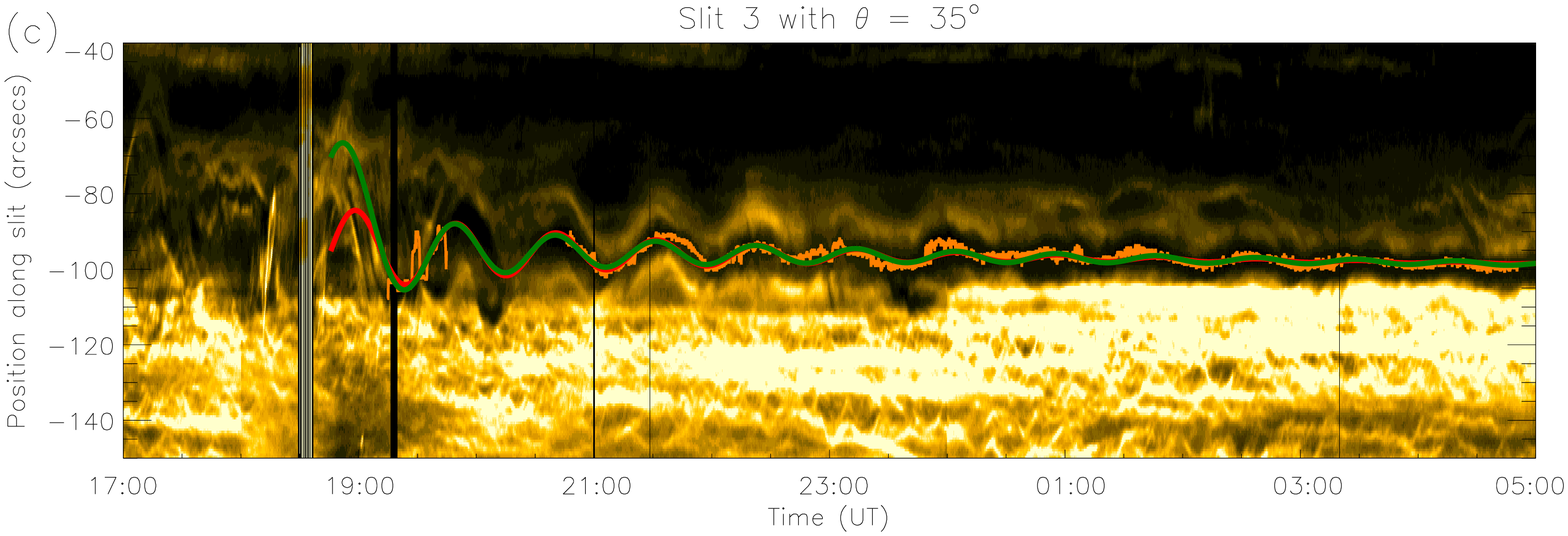}
\caption{Time-distance diagrams of the best slits at positions 1 (a), 2 (b), and 3 (c). The data of the positions of the center of the dark band are plotted as orange points. The fitted functions are also plotted as red and green lines corresponding to the exponentially decaying sinusoid and the Bessel function respectively. 
\label{fig:bestslits1}}
\end{figure*}

\begin{figure*}[ht]
\centering\includegraphics[width=18cm]{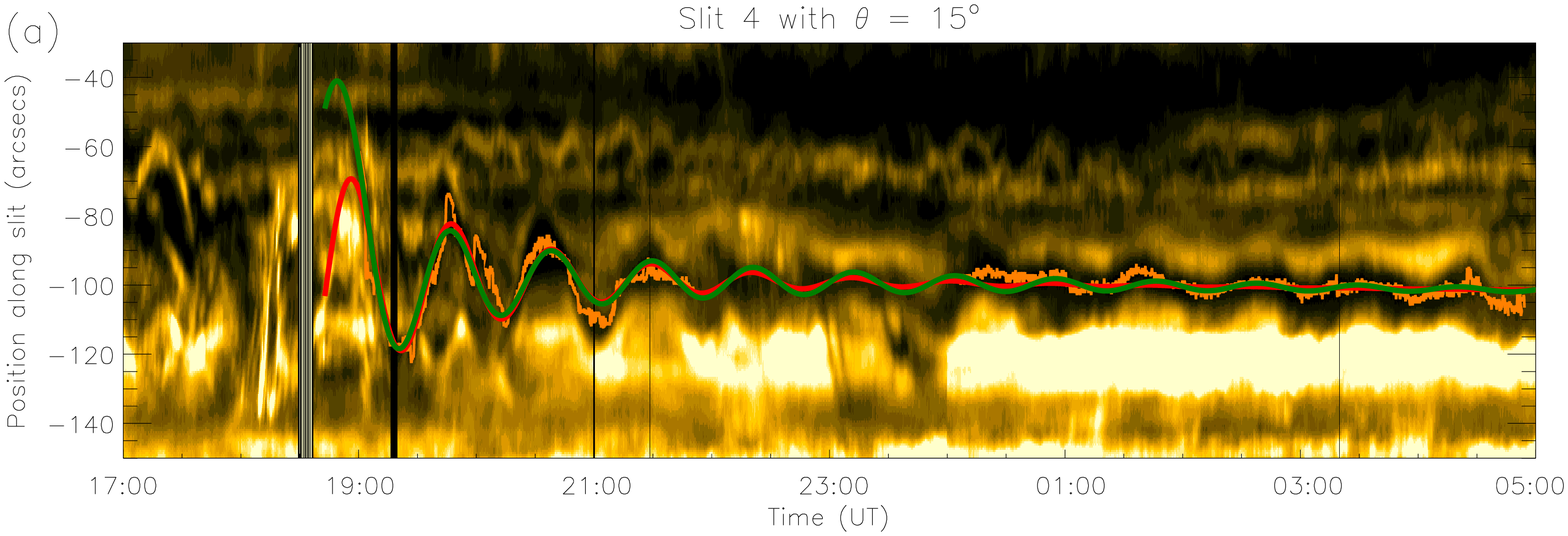}\\
\vspace{0.1cm}\centering\includegraphics[width=18cm]{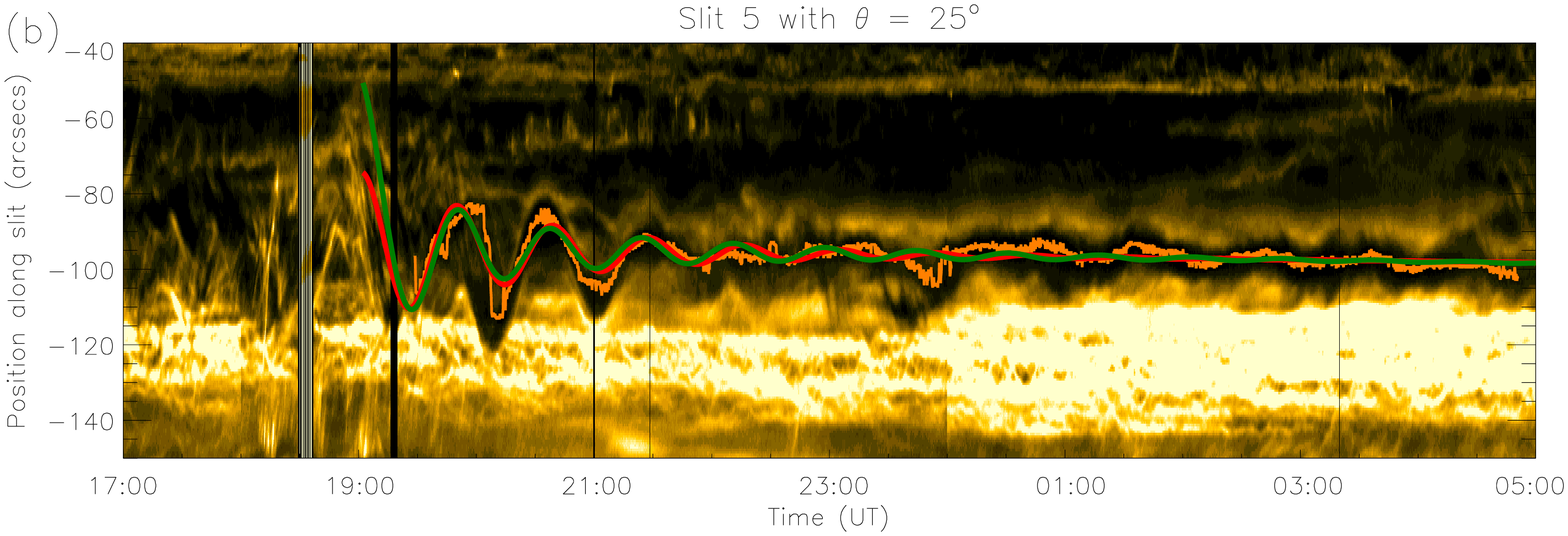}\\
\vspace{0.1cm}\centering\includegraphics[width=18cm]{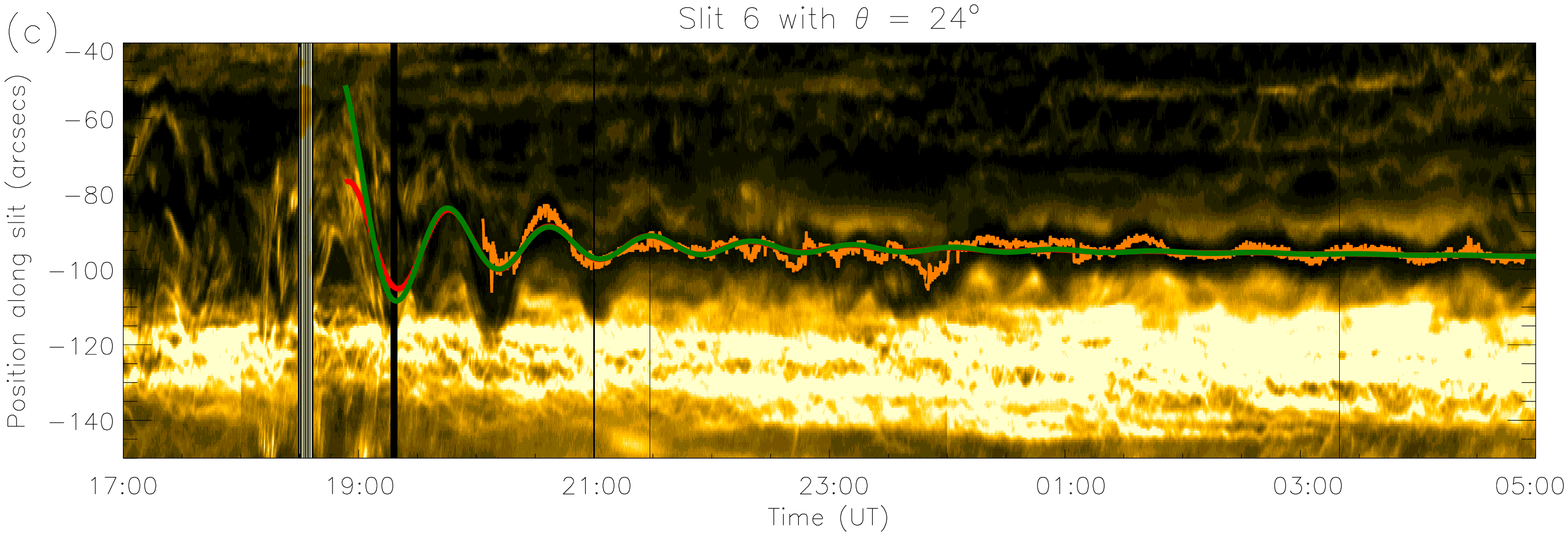}
\caption{As Fig. \ref{fig:bestslits1} for the slit positions 4 (a), 5 (b), and 6 (c). 
\label{fig:bestslits2}}
\end{figure*}

\begin{figure*}[ht]
\centering\includegraphics[width=18cm]{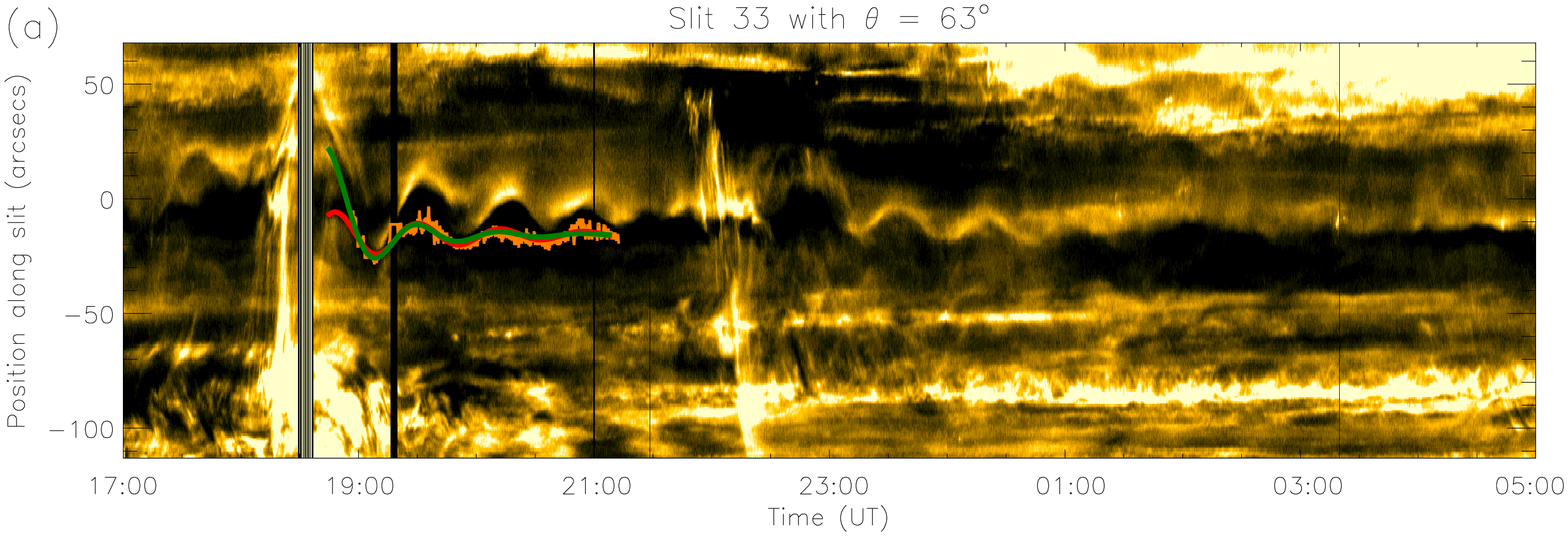}\\
\vspace{0.1cm}\centering\includegraphics[width=18cm]{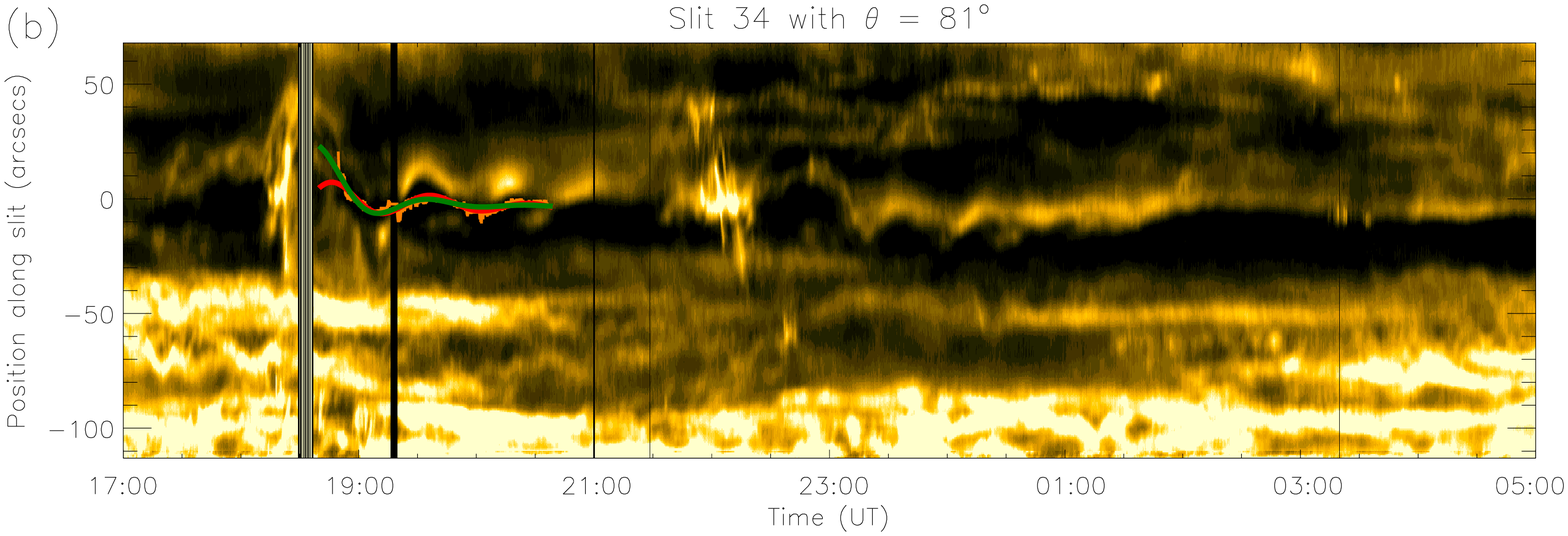}
\caption{As Fig. \ref{fig:bestslits1} for the slit positions 33 (a), and 34 (b). 
\label{fig:bestslits3}}
\end{figure*}

Figure \ref{fig:anglescan} shows 171 \AA\ time-distance diagrams for 3 slits centered at the same labeled position 1, differing only in the angle between the slit and the main direction of the filament. Although the oscillation is clear in all three panels, the oscillatory pattern in the time-distance diagrams depends strongly on this angle. In Figure \ref{fig:anglescan}(a), where the slit angle is $11^\circ$, several threads with different alignments contribute to the emission and the oscillation is only clear between $t=$ 20:00 and 22:00 UT.  Furthermore, this slit is not situated at the equilibrium position of the oscillation and is not aligned with the thread motion. Thus the threads enter and leave the slit as they oscillate, so only parts of the oscillation are seen in the panel. In Figure \ref{fig:anglescan}(b), the slit is tilted $21^\circ$ relative to the filament axis and oscillations are very clear from $t=$ 18:25 UT onward.  At this angle the dark band is seen throughout the observation, and the bright emission is symmetric around the cool plasma. Moreover, the dark band and adjacent bright emission oscillate together, indicating that the PCTR moves with the filament threads. In Figure \ref{fig:anglescan}(c) the slit angle is $31^\circ$ and the oscillations are clearly visible, but their amplitude is smaller than in Figure \ref{fig:anglescan}(b). Thus, at this angle the slit contains the equilibrium position of the oscillation, but the co-alignment with the actual motion of the threads is poorer than in Figure \ref{fig:anglescan}(b). Therefore we conclude that, of the three examples at position 1 along the filament, the $21^\circ$ slit is aligned best with the oscillating threads. 

This figure only compares the results of three different angles between the slit and the filament axis for one particular slit position. Overall we analyzed the oscillations by making time-distance diagrams for angles ranging from $0^\circ$ to $40^\circ$ at $1^\circ$ intervals, for the first 28 positions along the filament. For the remaining 8 positions it was necessary to increase the angular range to $0^\circ$ - $90^\circ$ in order to observe the oscillations. From the resulting 1840 time-distance diagrams, we selected the best slit direction for every position along the filament, by choosing the strongest oscillation pattern: a clear dark band sandwiched between 2 distinct bright emission regions, the longest possible time interval over which the oscillations were measurable, and the maximum possible displacement amplitude.

After creating time-distance diagrams for all 36 positions along the filament, we found that the filament only oscillated at positions 1-6, 33, and 34, while the remainder of the filament remained stationary. Figures \ref{fig:bestslits1}-\ref{fig:bestslits3} depict the best time-distance diagrams obtained with our method for the 8 oscillating positions. In Figure \ref{fig:bestslits1} the oscillation is very clear most of the time, but in the cases plotted in Figures \ref{fig:bestslits2} and \ref{fig:bestslits3} the oscillation is clear only between $t$ = 18:00 UT and 22:00 UT. Plasma from the nearby filament eruption (\S \ref{sec:observation}) blurs the oscillations in Figure \ref{fig:bestslits2}, but they reappear after $t$ = 23:00 UT. In all cases, a transient intensity increase related to the triggering appears around $t$ = 18:12 UT; then the threads start oscillating, reaching their maximum displacement by t $\sim$19:00 UT. This initial oscillation phase is very complex in all cases: several threads can be discerned as dark bands capped at both ends by thinner bright bands, but it is impossible to follow a clear and continuous dark band immediately after the maximum displacement has been reached. However, we were able to identify the initial oscillation by the bright emission. After the first half period a clear dark band oscillates but damps quickly, so the displacement has been reduced considerably within a few periods. It is important to note that, in the 6 cases shown in Figures \ref{fig:bestslits1} and \ref{fig:bestslits2}, we can distinguish oscillations nearly to the end of the temporal sequence. 

The movie of Figure \ref{fig:refmap} shows that part of the nearby erupted filament passes over all slit positions from the NW to the SE, taking one hour to travel across the field of view. This eruption produced visible brightening in slits 33 and 34 (Fig. \ref{fig:bestslits3}) at $t \sim$ 22:00 UT, and excited a new damped oscillation in the NW section of the filament. In Figures \ref{fig:bestslits1} and \ref{fig:bestslits2}, however, we see only cool (dark) plasma extending below the dark band during $t$ = 22:00 - 24:00 UT, and the  oscillations in the SE section of the filament were not disrupted.
 
We selected the position of the minimum emission intensity in the dark band at every time from the time-distance diagrams. For each of the 8 best slits, we selected an initial data point ($t_i$,$s_i$) along the dark band and found the coordinates ($t_i$,$s_{i,min}$) at which the intensity is minimized. We then moved to the next point in time ($t_{i+1}$, $s_{i,min}$) and scanned through the spatial range ($s_{i,min}-\epsilon$/2, $s_{i,min}+\epsilon$/2), where $\epsilon$ is of the order of the width of the dark region, to find the minimum-intensity location ($t_{i+1},s_{i+1,min}$). This process was repeated for the entire oscillation. Hence the coordinates of the filament thread are given by the set of points ($t_j,s_{j,min}$) with minimum intensity along the oscillating curve at times when the oscillation is clear. In Figures \ref{fig:bestslits1} - \ref{fig:bestslits3} the resulting ($t_j,s_{j,min}$) data are plotted as orange points. 

To characterize the oscillations quantitatively, we fit these data with two functions applicable to oscillating threads without and with mass accretion. For the constant-mass assumption we used an exponentially decreasing sinusoid of the form
\begin{equation}\label{eq:expfunc}
s(t)=s_0 + A_\mathrm{Exp} e^{-(t-t_0)/\tau} \cos \left[ \omega (t-t_0) +\phi_0 \right] +d_0 (t-t_0)~,
\end{equation}
where $s_0, t_0, \phi_0$, and $d_0$ are derived from the best fit. Here $s$ is a position along the slit, $t_0$ is the time of maximum displacement with respect to the oscillation center, $s_0$ is the oscillation center position at $t_0$, $\phi_0$ is the initial phase of the oscillation, and $d_0$ is the drift velocity. The central position as a function of time is then $s_0 + d_0 (t-t_0)$. This drift velocity accounts for slow displacements not associated with oscillations. However, in all fits $d_0$ is very small and the central position remains at $s_0$. The results of the fits of the exponentially decaying sinusoid are plotted in Figures \ref{fig:bestslits1} - \ref{fig:bestslits3} as red curves.

\citet{luna2012b} found that, if a filament thread continuously accretes mass, then its oscillation is described by a Bessel function instead of a sinusoid. Therefore we also fit the minimum intensity locations vs. time with
\begin{equation}\label{eq:besselfunc}
s(t) = s_0 + A_\mathrm{Bes} J_0\left[ \omega (t-t_0) + \psi_0 \right] e^{-(t-t_0)/\tau_w}+d_0 (t-t_0)~,
\end{equation}
where $\psi_0$ is related to the mass accretion rate (see \S\S \ref{sec:damp} and \ref{sec:accrete}) and $\tau_w$ is the weak damping time possibly associated with small, non-adiabatic energy losses. In Figures \ref{fig:bestslits1} - \ref{fig:bestslits3} the results of the Bessel function fits are shown by green lines.

The functions of Equations (\ref{eq:expfunc}) and (\ref{eq:besselfunc}) and the uncertainties in their parameters have been fitted to the data using IDL's CURVEFIT routine, which yields a nonlinear least-squares fit to a function of an arbitrary number of parameters. We also propagated errors to obtain the uncertainties of the derived quantities of the directly fitted parameters. 

Visually the two functional fits to the data are very good (see Figures  \ref{fig:bestslits1} - \ref{fig:bestslits3}). Both have the same number of free parameters, so their relative goodness can be assessed by comparing the standard deviation $\sigma^2$ of each function \citep[see][]{Asensio-Ramos2012}. Both fits have similar, reduced $\sigma^2$ values $\leq 13~\mathrm{arcsecs}^2$ for all 8 cases. Thus, the deviation of the data with respect to the fitted functions is less than 3.6 arcsecs, indicating that the fits are equally good. However, as demonstrated below, the modified Bessel function of Equation \ref{eq:besselfunc} provides better fits to the initial plasma dynamics and damping, and to the longevity of the oscillations.  

As noted earlier, the initial phase is very confusing and the dark band is not clear. Therefore we extrapolated the fitted functions from later times into this phase, and we tracked the behavior of the bright plasma to derive the initial thread displacements. We found that the modified Bessel function follows the initial motion of the bright plasma much better than the sinusoid-exponential; in particular, the predicted amplitude of the displacement is much closer to the observed values (see \S\ref{sec:displace}). In slit 1, for example, the first maximum displacement is $\sim$50 arcsecs with respect to the central position of the oscillation, whereas the displacement of the next maximum is $\sim$20 arcsecs, a 60 $\%$ reduction. In spite of this strong damping, the oscillations remain visible until the end of the temporal domain. Because the sinusoidal-exponential fit imposes the same damping at all times, this approximation cannot match both the initial strong damping and the long-term persistence of the oscillations.  In contrast, the modified Bessel fit can account for both the strong damping at the onset and the weak damping of the later oscillations. This conclusion applies equally to the other 7 slits, except for the interval where a secondary excitation occurs and a new damped oscillation is excited in slits 33 and 34 (see Figure \ref{fig:bestslits3}).

\section{Results}\label{sec:results}

\begin{figure*}
\includegraphics[width=9cm]{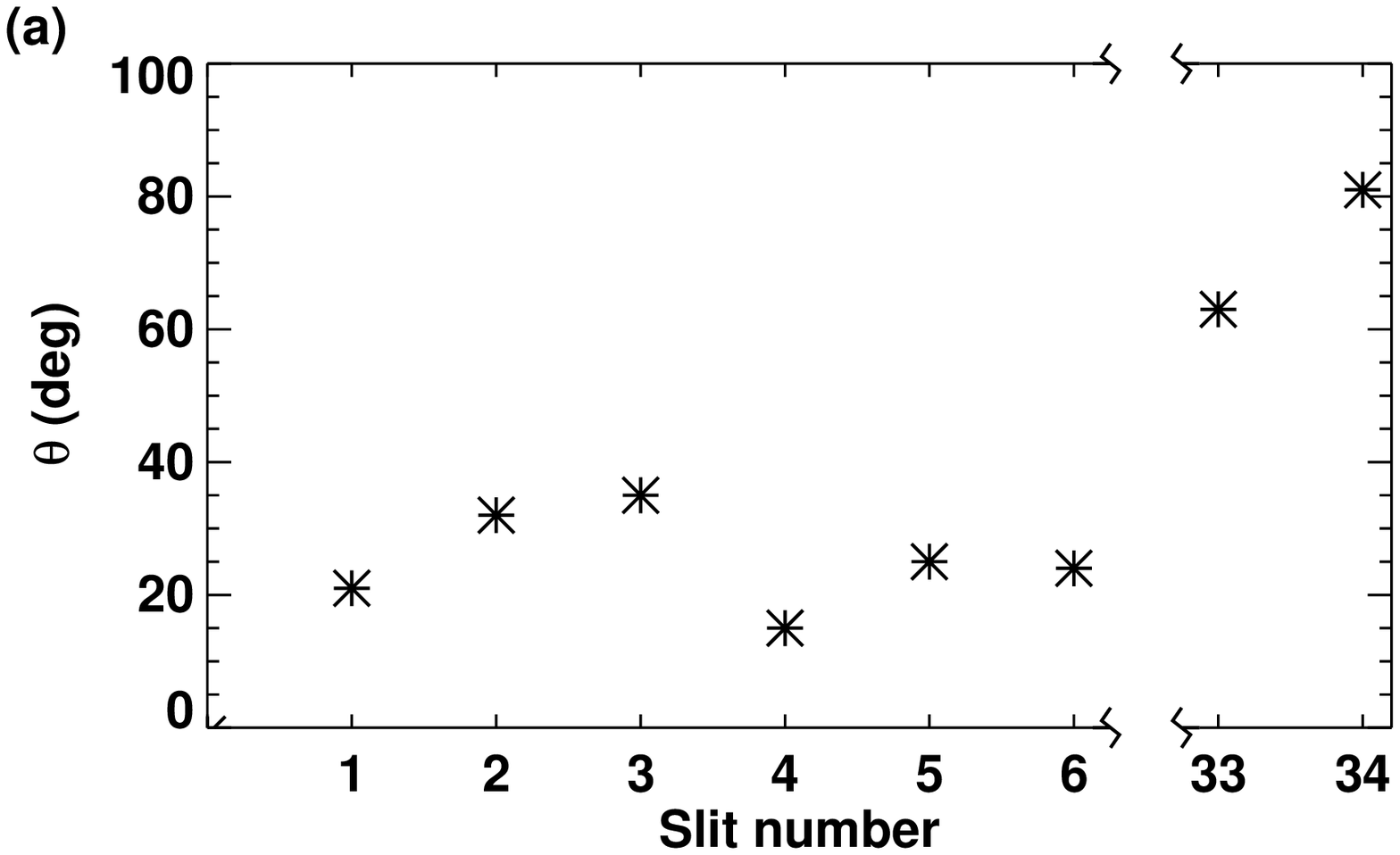}
\includegraphics[width=9cm]{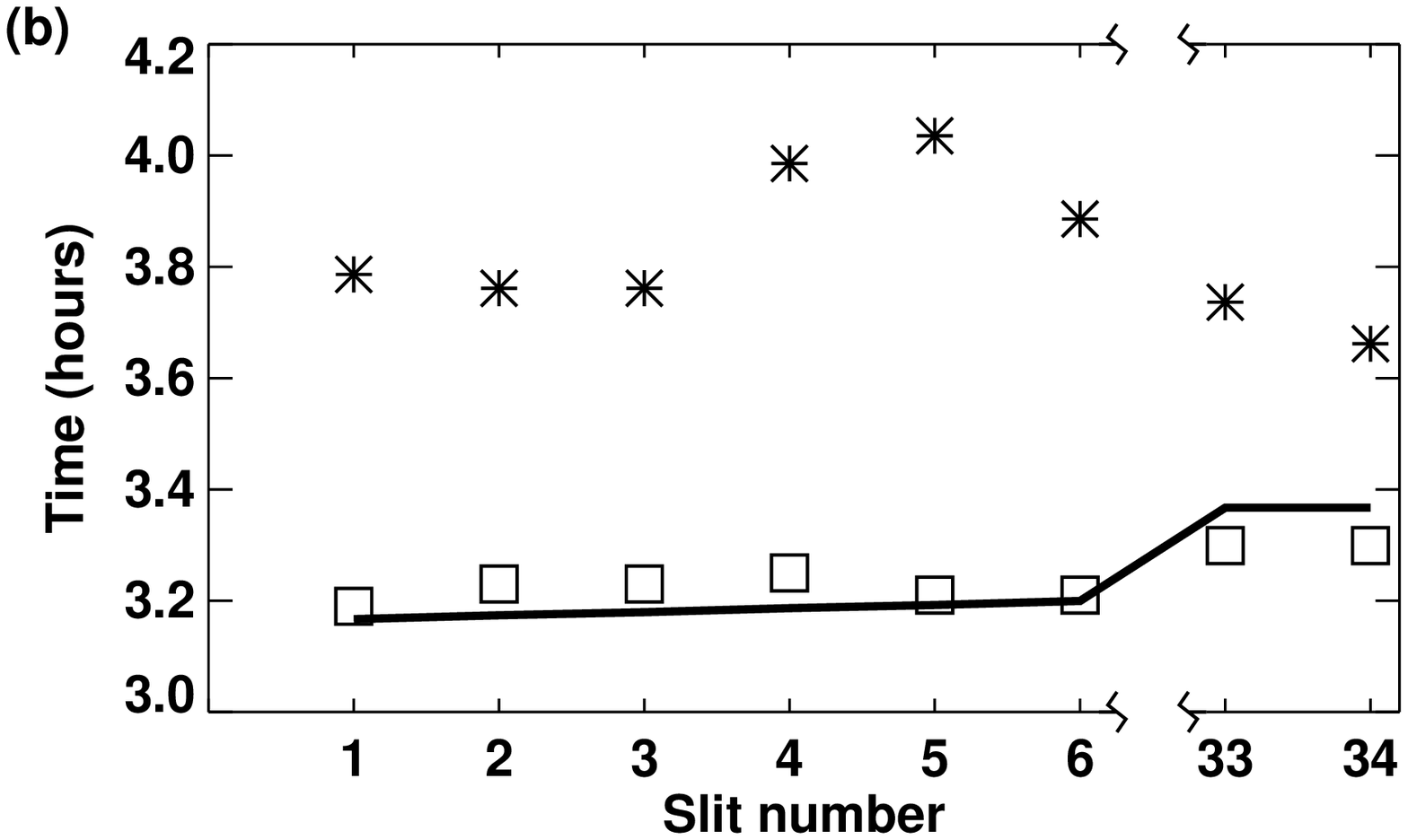}
\hspace{-0.5cm}\includegraphics[width=9cm]{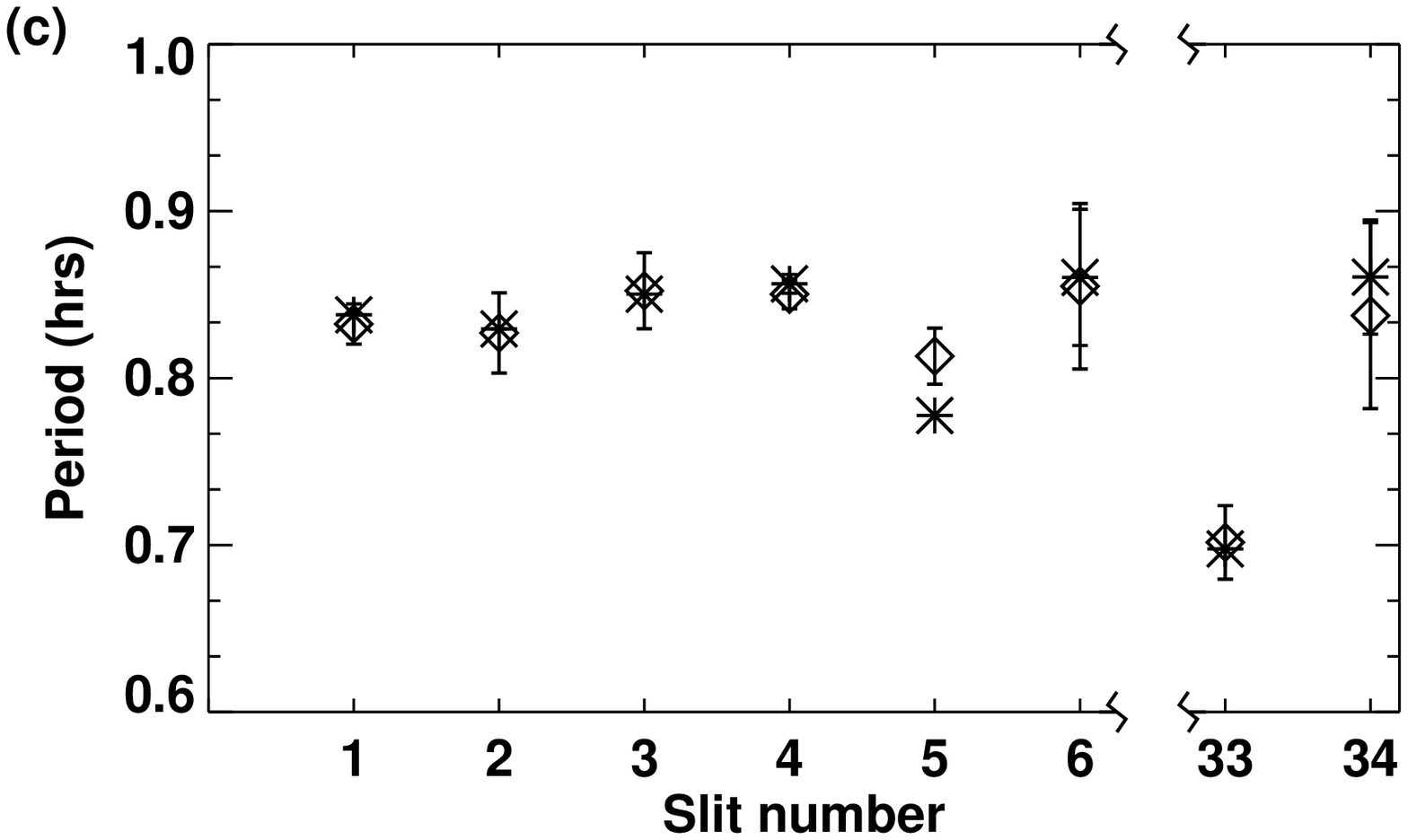}
\includegraphics[width=9cm]{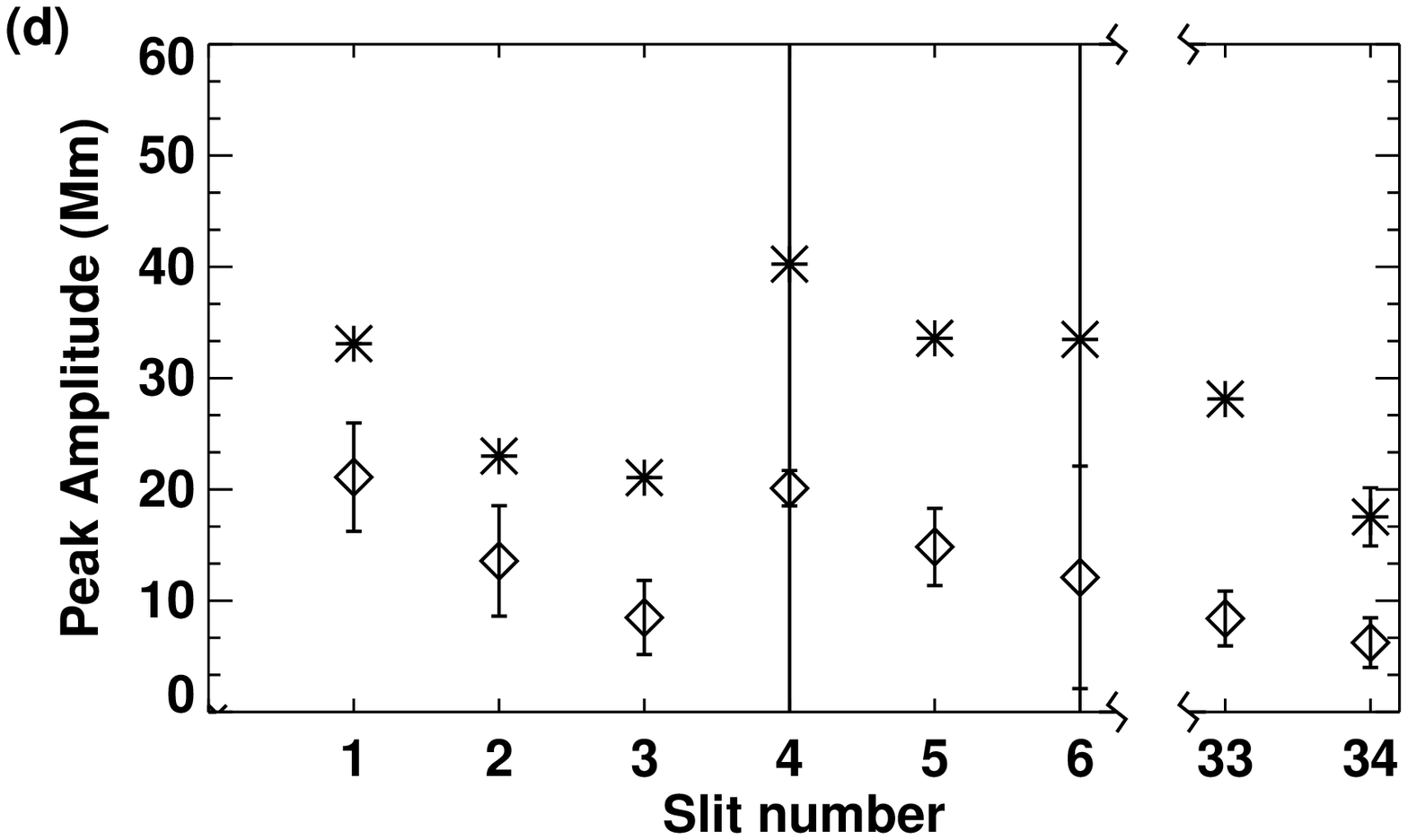}
\hspace{-0.5cm}\includegraphics[width=9cm]{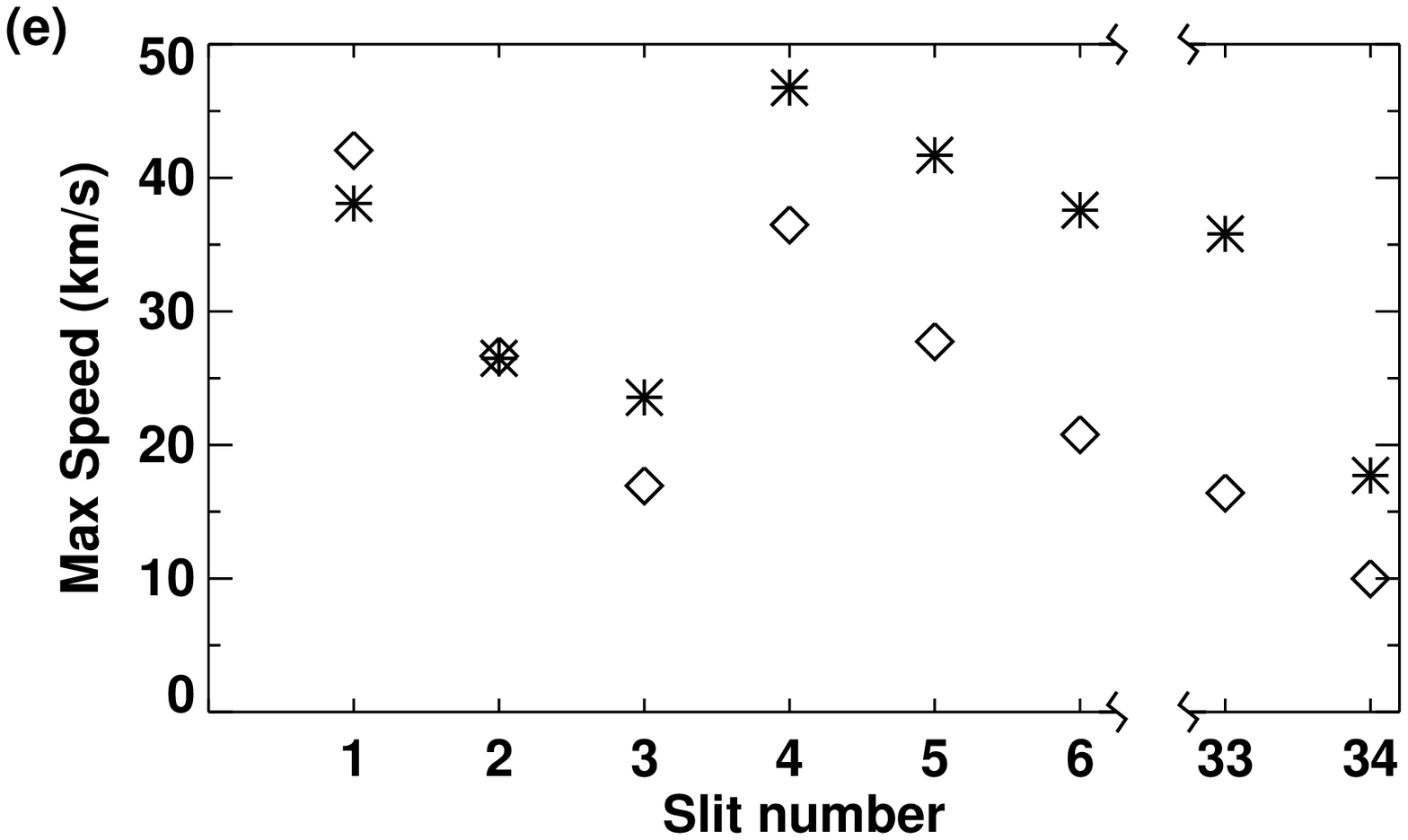}
\hspace{-0.1cm}\includegraphics[width=9cm]{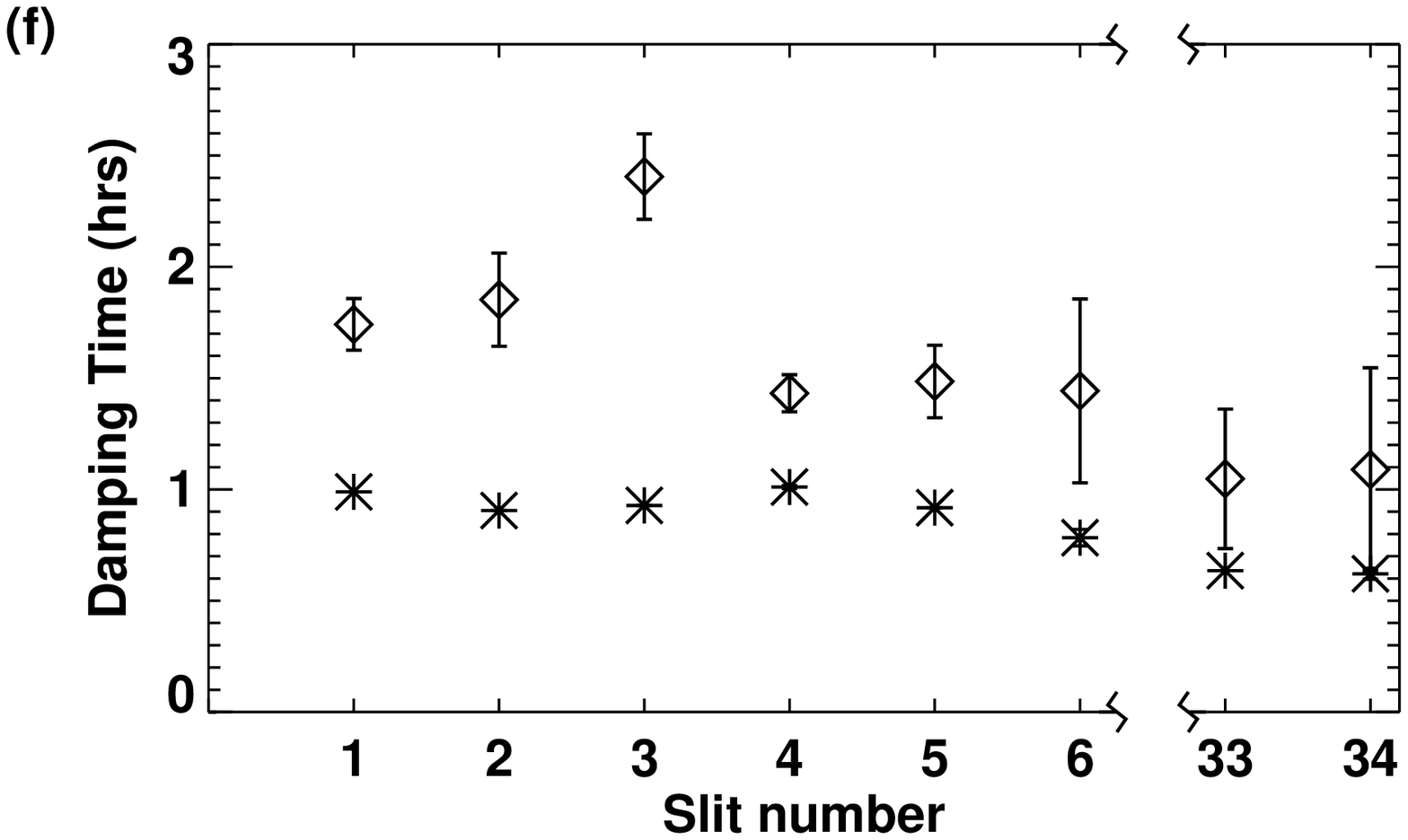}
\caption{Oscillation parameters for the 8 best slits. Where applicable, diamonds represent the exponential-sinusoid function fits (Eq. \ref{eq:expfunc}), and asterisks represent modified Bessel function fits (Eq. \ref{eq:besselfunc}). (a) Angle between the oscillation direction and the filament direction; (b) Times of maximum displacement ($t_0$; asterisks) and first 171\AA\ brightening (squares). Times of first brightening were estimated by eye. The solid line represents minimum travel times from the jet origin to the slit midpoints in straight trajectories, for a flow speed of $95~\mathrm{km~ s^{-1}}$. (c)  Period;  (d) Peak displacement; (e) Maximum speed; (f) Damping time (see text for details).  \label{fig:measured-parameters}}
\end{figure*}
\subsection{Oscillation direction}
With the method described in \S \ref{sec:analysis} we determined the main direction of oscillation at each slit position along the filament. As shown in Figure \ref{fig:measured-parameters}(a), the oscillation angles are in the range $\theta = 15^\circ-35^\circ$, with a mean value of $25^\circ$, in the first six positions corresponding to the SE end, and $\theta=63^\circ$ and $81^\circ$ in the last two positions, corresponding to the NW end. The change in angle between adjacent slit orientations was chosen to be $1^\circ$ for a given slit center position, so we consider the uncertainty of the angle to be $\pm 1^\circ$. The filament appears to contain ensembles of threads with similar orientations locally. Our method of measuring the orientation selects the thread or set of threads with the clearest oscillatory pattern. Therefore we expect that the orientation of the measured thread is applicable to all threads close to that slit. The angles estimated at the first six positions agree with published measurements of the magnetic-field orientation in prominences \citep[e.g.,][]{Leroy1983a, Leroy1984a,tandberg1995,trujillo2002,casini2003,Lopez-Ariste2006a}.

The filament structure undergoes an abrupt southward shift around slit 28, such that the NW segment of the filament apparently protrudes laterally from the PIL (see Figure \ref{fig:refmap}). The angles between the plasma motion in slits 33-34 and the main filament axis are $-3^\circ$ and $14^\circ$, indicating that the oscillating threads are almost parallel to the SE segment of the filament and not aligned with the protrusion. This suggests that the NW end of the filament might be a barb \citep{martin1998}, but it is difficult to determine from the available images and the HMI magnetograms whether this substructure is associated with a parasitic polarity or exhibits other definitive characteristics of a barb. To support the threads in their observed orientation, the magnetic field in this segment must be mostly horizontal, as proposed in some models of filament barbs \citep[e.g.,][]{aulanier1998,vanballegooijen2004,chae2005}. 

\subsection{Oscillation initiation}\label{sec:init}

The energetic event occurs at one edge of the filament channel containing the filament of interest, approximately 51 Mm from the closest visible thread end. The relative timing of the oscillation onset at different positions along the filament provides clues as to the mechanism whereby the impulsive energy release triggers and drives the oscillations. Two scenarios are possible: serial and parallel activation. In the ``serial" picture, a disturbance reaches the closest threads first, then propagates along the filament exciting sequentially the farther threads.
In the ``parallel" picture, the hot plasma travels simultaneously along many field lines within the filament channel, initiating oscillations when each stream within the jet reaches the thread supported by that field line. The SE footpoints of these field lines are connected to the source site, and the jet flows take longer to reach the more distant threads. Our analysis provides enough evidence to discriminate between these scenarios, and to select the most appropriate one. 

The onset times for the oscillations can indicate how the trigger reaches the threads. A good signature of this time of impact, in the time-distance diagrams, is the first observed brightening of the hot emission immediately south of the dark band (e.g., the cool thread), which occurs immediately before the oscillations start. As shown in Figure \ref{fig:measured-parameters}(b), the time of initial brightening increases slightly with slit position, indicating that the triggering first occurs closest to slit 1 and within a few minutes reaches more distant locations along the filament. However, a propagating disturbance moving from the SE to the NW part of the filament is not evident, as would be expected in the serial scenario. More importantly, the middle part of the filament remains at rest throughout the observation period; this is completely inconsistent with serial activation. The fact that we observe brightenings in the time-distance plots exactly where and when the threads begin oscillating strongly suggests that parallel activation is taking place. These brightenings occur when the hot jet flows reach the threads, and the oscillation starts exactly at this time. Close to the source the jet is highly collimated, but the larger-scale magnetic geometry guides this flow to cover a large fraction of the filament channel. Therefore we conclude that the jet excites the threads as predicted by the ``parallel'' picture. 

The directly fitted parameter $t_0$ indicates the time of maximum thread displacement, and thus should also be related to the oscillation onset time. However, $t_0$ depends on the velocity and mass of the threads at the moment when the oscillation was triggered. For example, filament plasma moving in the opposite direction to the trigger flow, due to counterstreaming \citep{zirker1998,alexander2013}, could delay the time of maximum displacement. In fact \ $t_0$ is earlier for slits 2, 3, 33, and 34 than at slit 1, which is closest to the energetic event. Thus, $t_0$ alone is not a good indicator of the time when the trigger reaches the threads.

We estimated a lower limit on the arrival time by computing the straight distance from the jet origin to the slit positions and dividing by the jet-flow speed. The arrival times (solid line) plotted in Figure \ref{fig:measured-parameters}(b) are computed with the maximum jet-flow speed of $95~\mathrm{km~ s^{-1}}$ determined in \S \ref{sec:trigger}. The derived and observed arrival times agree well, indicating that  the jet is responsible for the excitation of the thread oscillations. The last two positions do not fit the $95~\mathrm{km~ s^{-1}}$ prediction as well. One possible explanation is that small uncertainties in the determination of the flow speed and the path length will produce greater differences in the arrival times for the more distant positions. 

As shown in Fig. \ref{fig:frontspeeds}, each episode of jetting is slower than its predecessor. Our analysis of the relative timings also indicates that the first episode excites the oscillations, while the later episodes do not significantly perturb the already moving threads.

\subsection{Periods}

Both functional fits yield similar values for the oscillation period $P$ along each slit, ranging from 0.7 to 0.86 hours with a mean value of 0.82 hours. Furthermore the period is nearly uniform along the filament (Figure \ref{fig:measured-parameters}(c)), with the exception of slit 33, and agrees with previous observations of LAL oscillations. Although the oscillation period exhibits little variation, the oscillations are not completely in phase (see movie), indicating that the filament threads oscillate quasi-independently. Thus, the estimated periods reflect the local characteristics of the filament, which must be relatively uniform.

\subsection{Maximum Displacement and Speed}\label{sec:displace}

The range of displacement amplitudes for the exponential-sinusoid fit is 6 - 21 Mm, and for the modified Bessel fit from 17 to 40 Mm (Figure \ref{fig:measured-parameters}(d)). The Bessel function fit yields the largest amplitudes, and is more consistent with the observations during the initial phase, as discussed in \S \ref{sec:analysis}. The maximum speeds plotted in Figure \ref{fig:measured-parameters}(e) range between $10$ and $41~\mathrm{km~s^{-1}}$ for the exponential-sinusoid fit and from $17$ to $47~\mathrm{km~s^{-1}}$ for the Bessel fit. As with the periods, both fits yield similar speeds. The error bars of the amplitudes of the Bessel fits for slits 4 and 6 are very large, because the uncertainties are very sensitive to small changes in the data fitted. In these two cases the nominal displacement values are similar to the others, however, so the fit appears to be adequate. 

The maximum displacement and speed depend on the energy provided by the triggering event, which is located near the SE end of the filament (see \S \ref{sec:trigger}). Because the distance between the trigger site and the slit positions along the filament increases with slit number, one might expect the oscillation amplitude to decrease with slit number. However, we have not found any evidence of such a decrease. In fact the smallest speed is in position 3, close to the trigger, whereas the values at the far NW positions are similar to those at the SE positions.We speculate that the triggering flows follow the magnetic field lines that are confined in the filament channel. These field lines do not expand or diverge, much resulting relatively little attenuation of the jet kinetic energy.

\subsection{Damping times}\label{sec:damp}

The first three time-distance diagrams (Figure \ref{fig:bestslits1}) exhibit clear oscillations throughout the observing period, which can be fitted by the modified Bessel function alone (weak damping at later times). However, in slits 4-34 the quality of the data is reduced and the late oscillations are harder to measure. Although Figures \ref{fig:bestslits2} and \ref{fig:bestslits3} also show signs of thread motions throughout, there are fewer data points (orange dots) and poorer fits at later times because we could not identify a coherent oscillatory pattern then (see \S \ref{sec:analysis}). Therefore, the strong and weak damping components are most easily separable in the first 3 slits, yielding the most reliable estimates of the weak damping time ($\tau_w \sim 6-9$ hr). 

Independent of the quality of the Bessel function fit, the first period of the oscillation is very strongly damped. We computed the associated damping time of the first stage of the oscillation by adjusting a decaying exponential between the first peak at $t_0$ and the second peak, one period later at $t_0 + P$. In \S \ref{sec:analysis} we noted that the displacement is reduced by 60\% in the first period, which yields a damping time comparable to the oscillation period ($e^{-P/\tau} \approx e^{-1} \approx 0.4$). Figure \ref{fig:measured-parameters}(f) shows the resulting strong damping times at all 8 slit positions.  Thus, in the initial stage of the oscillation
\begin{equation}
\tau_\mathrm{strong} \sim P ~.
\end{equation} 
Figure \ref{fig:measured-parameters}(f) also shows the damping times of the sinusoid-exponential fit (Eq. \ref{eq:expfunc}), $\tau$, which range from 1.0 to 2.4 hours; the damping time per period is $\tau / P=$ 1.3-2.8. Because the exponential-sinusoid damping time is larger than the strong damping time of the modified Bessel function fit, suppressing the oscillations far too quickly, we conclude that the sinusoid-exponential fit cannot account for the initial strong damping. It is interesting to note that both damping times are almost uniform along the filament. The rapid damping implies that the associated damping mechanism must be very efficient for the LAL oscillations. We discuss this mechanism in terms of the thermal nonequilibrium model in \S \ref{sec:accrete}.

\subsection{Radius of curvature}

\begin{figure*}[!ht]
\vspace{-0.3cm}\centering\includegraphics[width=12cm]{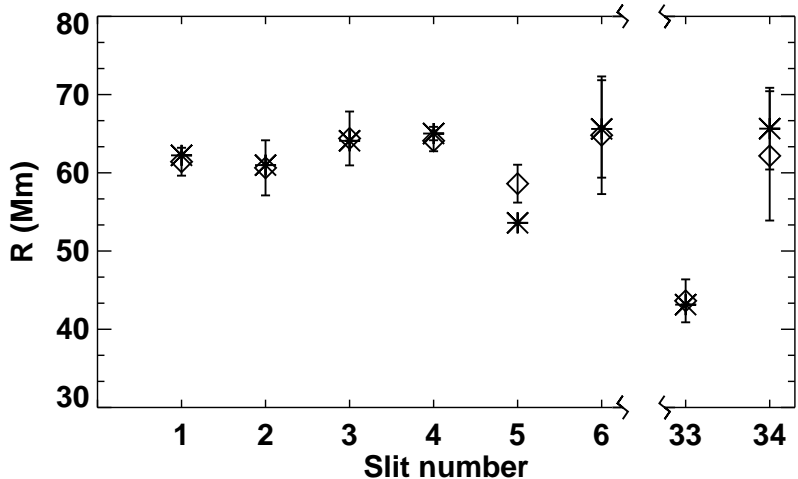}
\caption{Derived radius of curvature, $R$, of the dipped field lines supporting the threads (Eq. \ref{eq:pendulum}). Diamonds represent quantities derived from the exponential-sinusoidal function fits (Eq. \ref{eq:expfunc}), asterisks represent quantities derived from Bessel function fits (Eq. \ref{eq:besselfunc}).
\label{fig:inferred-parameters}}
\end{figure*}

Assuming that the restoring force is the projected gravity along the dipped field lines supporting the prominence plasma, the angular frequency of an oscillating thread is
\begin{equation}\label{eq:pendulum}
\omega =\frac{2 \pi}{P}=\sqrt{\frac{g_0}{R}}~,
\end{equation}
where $g_0$ is the solar gravity and $R$ is the radius of curvature of the dips \citep{luna2012b, luna2012c}. With this expression we obtained the radius of curvature of the dipped field lines of the observed filament using the calculated oscillation periods. In Figure \ref{fig:inferred-parameters} these radii are plotted for different positions along the filament; the range of values is $R= 43-66~\mathrm{Mm}$. The absence of any clear dependence of the magnitude on location suggests that the geometry of the field lines supporting the filament plasma was more or less uniform along the channel in the regions where oscillations were observed. 

\begin{figure*}[!ht]
\centering\includegraphics[width=12cm]{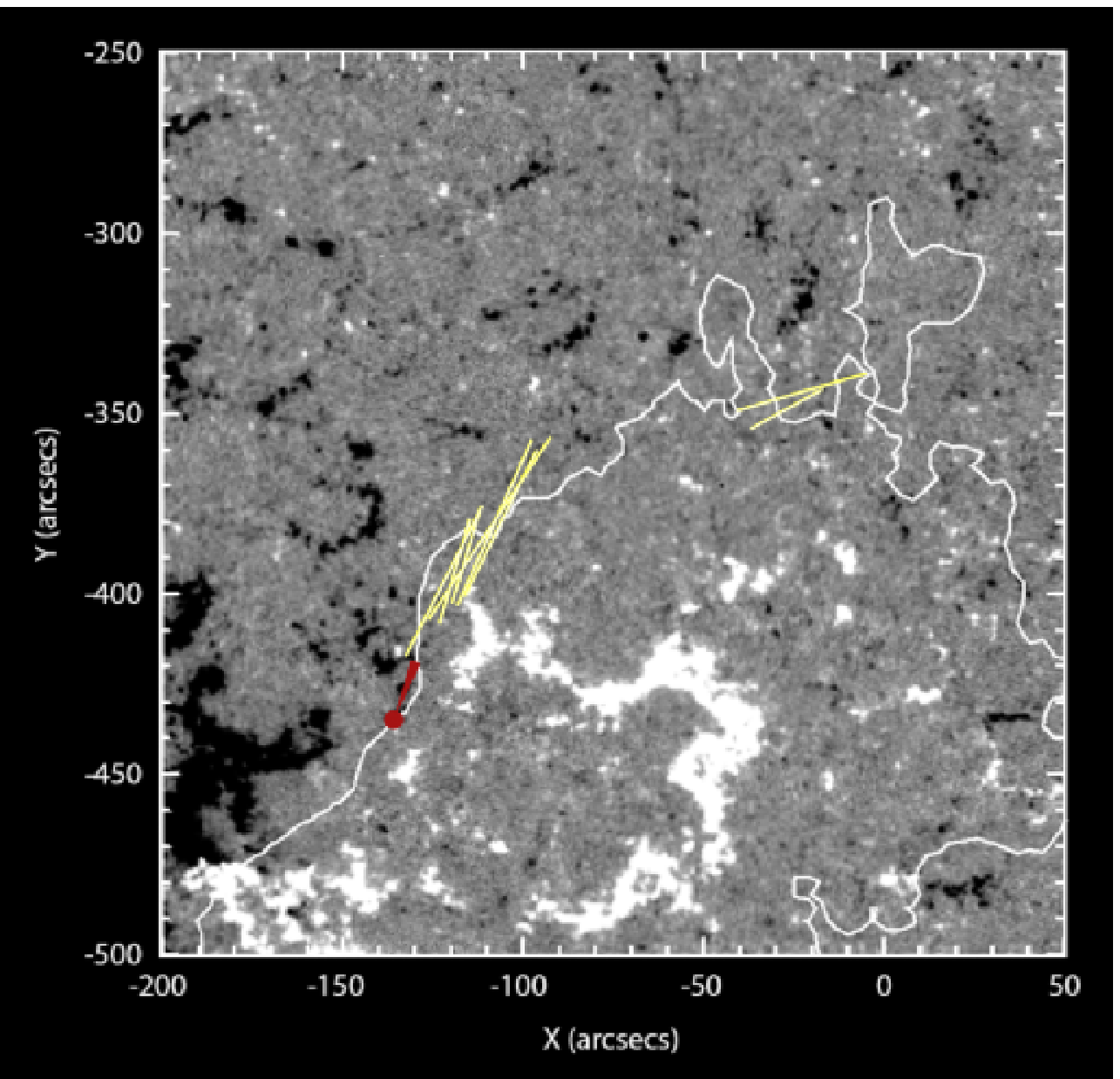}\vspace{0.3cm}
\centering\includegraphics[width=12cm]{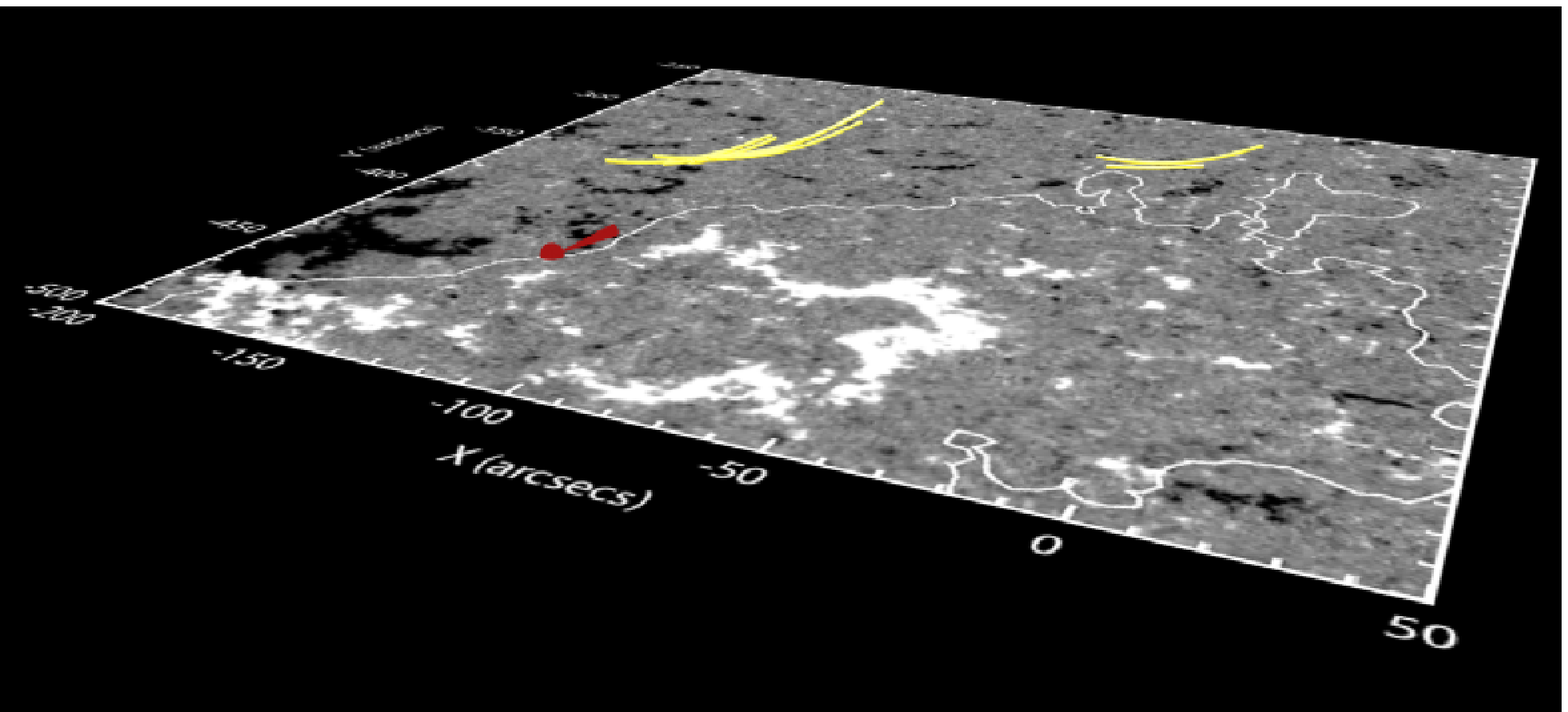}
\caption{The 3D geometry inferred from our analysis viewed from above (top) and the side (bottom). The grayscale plane shows a SDO/HMI magnetogram taken at 18:00 UT. The continuous white line over the magnetogram is the magnetic PIL of the region. The yellow lines represent the magnetic field lines supporting the filament plasma at the 8 best slit positions. Their radii of curvature correspond to the values of Fig. \ref{fig:inferred-parameters}. The red dome symbolizes the origin of the initiation jets and the red cone indicates the direction of the jet flows. All of the dipped field lines are plotted with the bottom of the dips 10 Mm above the photosphere; this height was arbitrarily chosen to yield a clear view of the dips on the bottom panel.
\label{fig:sketch}}
\end{figure*}

Figure \ref{fig:sketch} displays a 3D representation of the magnetic field structure inferred from our model at the 8 best slit positions, with a LOS HMI magnetogram as the background. The yellow curves representing the dipped parts of the field lines that support the oscillating threads are drawn with the orientations and radii of curvature given in Figures \ref{fig:measured-parameters}(a) and \ref{fig:inferred-parameters}. However, for clarity we placed the bottom positions of all dips 10 Mm above the photosphere. The threads oscillate around $s_0 + d_0~(t - t_0)$ (from Equations \ref{eq:expfunc} and \ref{eq:besselfunc}), but $d_0$ is very small or negligible according the function fits. Thus the threads oscillate around a fixed position, $s_0$. We also assumed that the bottom of the dips are the equilibrium positions of the thread displacements. These displacements are not symmetric with respect to the center position, as shown in Figures \ref{fig:bestslits1} - \ref{fig:bestslits3}: maximum elongation occurs on the side of the thread farthest from the jet, and only reaches approximately half of this amplitude on the same side as the jet due to the strong damping. The approximate position where the energetic event occurs (brightening) is indicated by a red dome, and the direction of the jet as a red cone.  

\subsection{Minimum magnetic-field strength}\label{sec:field}

We computed the minimum magnetic-field strength at the field line dips, assuming that the magnetic tension should at least balance the weight of the threads. Previously we found the following expression for the minimum value of the magnetic field as function of the electron number density $n_\mathrm{e}$ \citep{luna2012b}:
\begin{equation}\label{eq:magfield}
B[\mathrm{\mathrm{G}}] \ge 26\, \left(\frac{n_\mathrm{e}}{10^{11}~\mathrm{cm}^{-3}}\right)^{1/2}\,P[\mathrm{hours}]~.
\end{equation}
The electron density in this filament was not measured, so we assume typical values in the range $10^{10} -10^{11}~\mathrm{cm^{-3}}$ \citep[see, e.g.,][]{labrosse2010}. We consider the electron density to be the most important source of uncertainty in the estimated magnetic field, because the uncertainties associated with the fit are smaller than the above density range. With this consideration Equation \ref{eq:magfield} yields
\begin{equation}\label{eq:magfield-corrected}
B[\mathrm{\mathrm{G}}] \ge \left( 17 \pm 9 \right) \,P[\mathrm{hours}]~.
\end{equation}
Because the oscillation period is nearly uniform along the filament, the variation of the magnetic field along the filament is smaller than the uncertainty associated with the density. Thus, the minimum magnetic field strength along the filament is $14 \pm 8~\mathrm{G}$, which is consistent with typical directly measured field strengths \citep[see, e.g.,][]{mackay2010}.

\subsection{Mass accretion rate}\label{sec:accrete}

In our model \citep{luna2012b}, the LAL oscillations are damped by continuous mass accretion onto the filament threads at rate $\alpha$, and the oscillations are described by the Bessel function of Equation (\ref{eq:besselfunc}). The phase $\psi_0$ is related to the mass accretion rate as $\psi_0=\omega m_0/\alpha$, where $m_0$ is the mass of the thread at $t_0$ and $\psi_0$ and $\omega$ are derived from the Bessel function fit. To obtain $m_0$, we assume that the thread is a cylinder of length $l$, radius $r$, and uniform electron density $n_\mathrm{e}$. Thus, $m_0=1.27 m_\mathrm{p} n_\mathrm{e} \pi r^2 l$, where $1.27 m_\mathrm{p}$ is the average coronal particle mass and $m_\mathrm{p}$ is the proton mass. Taking a typical thread radius $r=100~\mathrm{km}$ \citep[e.g.,][]{lin2003} we obtain
\begin{equation}\label{eq:threadmass}
m_0 \mathrm{(kg)} = 6.67 \times 10^6~\left(\frac{n_\mathrm{e}}{10^{11}~\mathrm{cm^{-3}}}\right) ~l \mathrm{(Mm)}~.
\end{equation}
The thread length is estimated to be the slit-aligned length of the central dark (absorption) region of Figures \ref{fig:bestslits1} - \ref{fig:bestslits3}, measured at the second period after the maximum displacement time in all cases. This procedure overestimates the real length of the thread because we are not considering the PCTR at both ends of each thread (which should be thin), and ignoring the fact that the 171 \AA\ dark region probably contains several threads along the LOS.  As for the minimum magnetic-field strength, the mass $m_0$ depends on an assumed electron density range. Thus
\begin{equation}\label{eq:threadmass_corrected}
m_0=\left( 4 \pm 3 \right) \times 10^6~l \mathrm{(Mm)}~.
\end{equation}
The thread length, $l$, is quite uniform along the filament, so the variation of $m_0$ along the filament is smaller than its uncertainty. With this mass estimate and the measurement of $l$, the mass accretion rate is estimated to be
\begin{equation}
\alpha=\left(36 \pm 27 \right) ~\times 10^6 ~\mathrm{kg ~hr^{-1}}~,
\end{equation}
which is consistent with the theoretical values predicted by our thermal nonequilibrium model with steady heating \citep[e.g., ][]{karpen2006,luna2012a}. The theoretical accretion rates are slightly smaller than the observed values, primarily because the present observational study has significant uncertainties in the thread length, radius, and density. In addition, these differences could be associated with small temporal and spatial variations in the heating at the footpoints. Consequently we consider the factor-of-2 agreement between the observed and predicted mass accretion rates to be quite good.

 \subsection{Jet energy estimate}\label{sec:energy}

As discussed in \S \ref{sec:trigger}, the filament oscillations are triggered by a small jet close to the southeast end of the filament. A highly collimated flow of hot plasma emanates from this small area at a projected speed of $\sim95~\mathrm{km~s^{-1}}$, yielding oscillating filament plasma that reaches speeds of $\sim 50$ km s$^{-1}$. We estimate the energy released to the filament by the jet event by computing the kinetic energy of the oscillations: $E=1/2 M_\mathrm{osc} v^2$, where $M_\mathrm{osc}$ is the oscillating thread mass and $v$ is the averaged velocity amplitude of the oscillation ($\sim30~\mathrm{km~s^{-1}}$). Because only 8 of 36 slit positions exhibit periodic motions, we estimate that only $8/36$ ($\sim$ 22\%) of the total filament mass oscillates. With these considerations the energy imparted by the jet to the filament oscillations is
\begin{equation}
E=10^{26}\left(\frac{M}{10^{14} g}\right)~\mathrm{erg}~.
\end{equation}
For a typical prominence mass $M=10^{12}-10^{15} ~ g$ \citep{labrosse2010}, the energy of the oscillating filament is predicted to be $E=10^{24}-10^{27}~\mathrm{erg}$. The energy transferred to the filament by the jet is an unknown fraction of the total energy released by the initiating event, however, so $E$ places a lower limit on the trigger energy.  \cite{zhang2013a} carried out numerical experiments generating LAL oscillations by impulsive heating at one footpoint of a loop, and found that the energy of the thread oscillations is only  4\% of the impulsive energy release. In our case the initiating event manifests several key characteristics of a microflare, most notably the presence of a jet, the duration, and the energy range \citep[see reviews by][]{benz2008,shibata2011}.

\section{Filament structure}\label{sec:structure}

\begin{figure}
\centering\includegraphics[width=7.5cm]{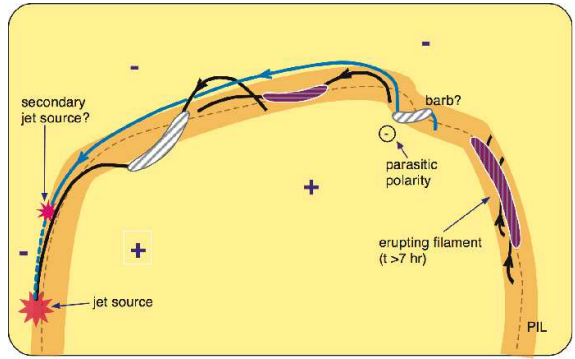}
\caption{Schematic picture of the magnetic connections between the jet (red star) and selected filament threads (striped regions showing the thread directions). The PIL is indicated by a dashed line; the broad orange ribbon is the filament channel. Field lines (black and blue solid lines) supporting white striped areas carry jet plasma from the energetic event to the oscillating threads, while purple striped areas represent threads along field lines that are not connected to the jet and hence do not oscillate.\label{fig:cartoon}}
\end{figure}

Our results provide important clues about the structure of the entire filament channel, which is sketched in Figure \ref{fig:cartoon}. As shown in \S \ref{sec:observation} the filament channel is curved, forming a semi-circular structure. In order to reproduce the magnetic field direction at the slits and the hemispheric rules, the chirality of the filament should be sinistral, in agreement with \citet{Wang2013a} who found that the barbs are left bearing. The NW substructure traversed by slits 28-36 is most likely a barb pointing to a small, negative parasitic polarity within the mainly positive region.  The PIL in this part of the active region is highly irregular, so it is difficult to discern the main direction of the PIL there. However, a more detailed study of this feature is outside the scope of this paper. The cartoon of Figure \ref{fig:cartoon} is consistent with the 3D reconstructed field of Figure \ref{fig:sketch}, in terms of the orientation and location of the dipped field lines in the oscillating segments.

At the western side of the channel, Figure \ref{fig:cartoon} depicts a region not analyzed in this paper: a filament segment that erupts around 21:40 UT. The oscillations in slits 33-34 are disrupted by the eruption, whereas the oscillations in slits 1-6 are largely unaffected. Therefore we expect that the field lines supporting the barb are linked to this erupting region, while the field lines supporting the SE part of the filament (slits 1-6) probably are not connected to the erupting filament segment.

The position of the jet is crucial because it determines the magnetic field connectivity to the oscillating threads. We assume that the field lines supporting these threads are magnetically connected at one footpoint with the jet source. The source position appears to be very close to the PIL, but the inherent uncertainty in the magnetograms makes it difficult to determine whether the jet site is north or south of the PIL. Given the southern location of the filament, the filament chirality, the predominance of inverse polarity in filaments, and the observed direction of the threads, however, we conclude that the jet source is most likely situated north of the PIL where the background field is predominantly negative.

\section{Summary and Conclusions}\label{sec:conclusion}

In this work we have studied the LAL oscillations of a filament and the associated trigger, using observations and theory to determine key properties of the filament and trigger. The LAL oscillations are displacements of low $\beta$ plasma supported by the magnetic field, so the direction of motion reflects the direction of the local magnetic field. Our results agree with the few direct measurements of the orientation of the magnetic field in filament threads with respect to the associated PIL. Thus, the oscillation analysis described in this work is a novel tool to determine the orientation of the filament magnetic field. We determined fundamental characteristics of the LAL oscillations and the triggering flows by fitting curves to the SDO/AIA time-distance diagrams. We used a exponentially decaying sinusoid and a modified Bessel function to fit the oscillation data, respectively representing constant-mass and mass-accreting solutions. Both fits generally match the data well, but the Bessel function fits the initial stage of the oscillation significantly better. Therefore mass accretion is likely to play a major role in damping the oscillations rapidly. We conclude that our earlier model for the damped LAL oscillations \citep{luna2012b, luna2012c} accurately explains the behavior of this filament. 

Using our earlier analytic approximations, we determined the radius of curvature and minimum strength of the magnetic field lines that support the filament plasma, and inferred the magnetic structure of the oscillating portions of the filament (Figures \ref{fig:sketch} and \ref{fig:cartoon}). We found that the geometry varies little along the filament, demonstrating that the different parts of the filament form a quasi-coherent structure whose origin and subsequent evolution remain linked. The resulting structure, as well as the presence of LAL oscillations, are compatible with the two leading magnetic-structure models --- the flux rope and the sheared arcade --- which predict that the bulk of the filament plasma resides in the dips \citep{mackay2010}. Although both models predict dipped field lines that can host oscillating threads, the distribution of cool mass condensed in these dips through thermal nonequilibrium has been studied in depth only for the sheared arcade model.

The thermal nonequilibrium model for filament mass formation predicts that the existing threads accrete material at the same rate as the chromospheric evaporation rate, as long as the standard coronal heating is localized at the footpoints. We established previously that this continuous accretion of mass is responsible for the strong damping of the LAL oscillations \citep{luna2012b}. The mass accretion rate of the filament threads computed in the present study agrees with the predictions of our thermal nonequilibrium model \citep{karpen2006, luna2012c}, and hence with the well-established quiet-Sun coronal heating rate of \cite{withbroe1977}. Based on our results, we suggest that LAL oscillations provide a new opportunity for constraining coronal heating models beyond the usual analyses of AR coronal loops.  

We propose the following general picture of the event and the filament structure. A reconnection process takes place close to PIL at the northern side. The resulting jet plasma flows along the filament channel field lines at a projected speed of $\sim95~\mathrm{km~s^{-1}}$. These field lines only connect with some parts of the filament, such that the flow reaches the SE part and the NW barb. Threads in these regions are pushed by the hot flows, then oscillate in the dips with a motion resembling a pendulum. Other complex physical phenomena could take place when the hot flows reach the cool plasma, but a detailed study of this interaction is beyond the scope of this work. The restoring force of the oscillations is the projected gravity along the dips, as our model predicts.  Continuous, localized coronal heating produces evaporation of chromospheric plasma that accretes onto the already formed filament threads. This mass accretion is responsible for the initial strong damping of the oscillations.

More observations of LAL oscillations in filaments and the associated triggering events, together with detailed simulations of the response of filament threads to hot flows, are needed to improve our understanding of this intriguing phenomenon. Additional theoretical modeling of LAL oscillations also will advance the use of seismology to probe the ambient physical conditions in filaments, as demonstrated here. Further analyses of LAL oscillations would benefit as well from greater understanding of the coronal heating mechanism, as the likely driver of mass accretion onto filament threads. We anticipate significant progress on these questions to be made in the near future by the combined capabilities of SDO, IRIS, and the upcoming Solar Orbiter mission.

\acknowledgements

ML gratefully acknowledge partial financial support by the Spanish Ministry of Economy through projects AYA2011-24808 and CSD2007-00050. This work contributes to the deliverables identified in FP7 European Research Council grant agreement 277829, ``Magnetic Connectivity through the Solar Partially Ionized Atmosphere", whose PI is E. Khomenko. KK acknowledges support for this work by a co-op agreement between the Catholic University of America and NASA Goddard Space Flight Center, sponsored by NASA's Heliophysics LWS and SR programs. KM gratefully acknowledges funding from the National Science Foundation
via grant \# 0962619. SDO is a mission for NASA's Living With a Star program. HG, JK, and TK also thank the LWS TR\&T Program for support. We also thank S. Antiochos, I. Arregui, A. Asensio-Ramos, J. L. Ballester, C. R. DeVore, A. D\'{\i}az, A. Lopez-Ariste, F. Moreno-Insertis, R. Oliver, D. Orozco-Suarez, and J. Terradas for helpful discussions and suggestions.

\end{document}